# An Optimal Control Model of transmission dynamics of COVID-19 in Ghana

Joseph Ackora - Prah[1], Samuel Okyere[1*], Ebenezer Bonyah[2], Kwaku Forkuoh Darkwah[1]

[1]Department of Mathematics, Kwame Nkrumah University of Science and Technology, Kumasi, Ghana

[2]Department of Mathematics Education, Akenten Appiah-Menka University of Skills Training and Enterpreneurial Development, Kumasi, Ghana

Corresponding Author: Samuel Okyere; okyere2015@gmail.com

**Abstract:**

Almost every country in the world is battling to limit the spread of COVID-19. As the world strives to get an effective medication to control the disease, appropriate intervention measures, for now, remains one of the effective methods to reduce the spread of the disease. Optimal control strategies have proven to be an effective method in curtailing the spread of infectious diseases. In this paper, a model has been formulated to study transmission dynamics of the disease. Basic properties of the model such as the basic reproduction number, equilibrium points and stability of the equilibrium points have been determined. Sensitivity analysis was carried on to determine the impact of the model parameters on the basic reproduction number. We also introduced a compartment for the deceased and examined the behaviour of COVID-19 related deaths. The numerical simulation prediction is consistent with real data from Ghana for the period March 2020 to March 2021. The simulation revealed the disease had less impact on the population during the first seven months of the outbreak.To help contain the spread of the disease, time dependent optimal controls were incorporated into the model and Pontryagin maximum principle was used to characterize vital conditions of the optimal control model. Numerical simulations of the optimal control model showed that, combination of optimal preventive strategies such as nose mask and vaccination are effective to significantly decrease the number of COVID-19 cases in different compartments of the model. Vaccination decreases the susceptibility to the disease whereas mask usage preserved the susceptible population from extinction.

**Keywords:** SARS-CoV-2, COVID-19, Stability analysis, optimal control, reproductive number, vaccination.

## 1 Introduction

The new pandemic is the SARS – CoV- 2 (COVID-19), which originated in Wuhan, China, in December 2019 and has spread through the entire globe [7]. Countries affected by the disease faced several waves of the pandemic which led to economic recessions with its implications for which Ghana is no exception. The country is now highly indebted due to various policies undertaken by the Government in the course of fighting the COVID-19 [27], spreads through the air, primarily via small droplets or particles such as aerosols, produced after an infected person breathes, coughs, sneezes, and talks [5, 15]. Symptoms include fever [11], cough, headache [10], fatigue, breathing difficulties, and loss of smell and taste [1, 8]. The symptoms may begin usually from 1 – 14 days after exposure to the virus. At least a third of the people who are infected become asymptomatic [4]. Of those who develop symptoms, 81% are mild to moderate , while 14% are severe and 5% suffer critical conditions [14].

Globally, over 500 million people worldwide have contracted the disease with over 6.2 million deaths. African countries have so far reported over 11.8 million cases and 253,422 deaths. Ghana confirmed its first case on the 12$^{th}$ of March 2020 when two infected people came to Ghana, one from Norway and the other from Turkey [25]. As of 19th April 2022, the country had reported 161,101 confirmed cases [18] with the majority of the cases from Accra and Kumasi.

The Government instituted a lot of measures to prevent more cases including a lockdown imposed on Greater Accra and Greater Kumasi, a ban on all social gatherings, and a closure of air and land borders. The lockdown restrictions on these two cities were lifted on the 20th of April, 2020 and after that, the mandatory wearing of nose masks, hand washing with soap and water, ventilating indoor spaces and the use of sanitizers were new measures imposed on Ghanaians by the government. Notwithstanding all these protocols, the government has introduced vaccination, and the Health Authorities are vaccinating all individuals aged 18 years and above [28].

The transmission dynamics of infectious diseases have been studied and analyzed by researchers using mathematical models. Models offer a simplified representation of reality and may be used to predict future outcomes of diseases and possible interventions.  COVID -19 has been studied [2, 6, 12, 13, 29 - 33], but since many aspects related to the COVID-19 virus are unknown, researchers continue to propose models that will best describe the dynamics of the disease. Ndaïrou et al. 2020 [29], proposed a compartmental mathematical model for the spread of the disease with a special focus on the transmissibility of super-spreaders individuals in Wuhan China. Mugisha et al. 2021 [30], proposed a mathematical model that incorporates the currently known disease characteristics and tracks various intervention measures in Uganda. They modeled the trace-and-isolate protocol in which some of the latently infected individuals tested positive while in strict isolation. Garba et al. (2020) [31] proposed a compartmental model to analyze the transmission dynamics of the disease in South Africa. A notable feature of their method was the incorporation of the role of environmental contamination by COVID-infected individuals. Fu et al. (2020) [32], applied Boltzmann-function-based regression analyses to estimate the number of SARS-CoV-2 confirmed cases in China. Nana-Kyere et al. 2022 [33], proposed the SEQIAHR compartmental model of the disease to provide insight into the dynamics of the disease by underlining tailored strategies designed to minimize the pandemic. Dwomoh et al. (2021) [34] proposed the SEIQHRS (susceptible-exposed-infectious-quarantine-

hospitalized-recovered-susceptible) model that predicts the trajectory of the epidemic to help plan an effective control strategy for COVID-19 in Ghana using the generalized growth model.

Optimal control methods applied in various models have contributed greatly in engineering intervention strategies in curtailing the spread of infectious diseases [37, 40]. Optimal control methods have been used to model diseases such as tuberculosis [40 - 42], HIV/AIDS [43], diabetes [44] and recently COVID-19 [45 - 49]. In [49], a mathematical model to study transmission mechanism in Senegal incorporating vital dynamics of the disease and two key therapeutic measures: vaccination of susceptible individuals and recovery/treatment of infected individuals was formulated. Nana – Kyere et al. (2020) [45], used an SEQIAHR compartmental model to study the disease dynamics and modified the model incorporating optimal controls such as personal protection and vaccination of the susceptible individuals. In [47], a model was used to study the disease transmission in Ethiopia. The model was modified incorporating optimal control strategies such as public health education, personal protective measures and treatment of hospitalized cases.

Motivated by the available COVID-19 works in all these literature, we seek to formulate an optimal control model which accurately predicts and suggest intervention strategies to curtail the spread of the disease in Ghana. The model considers two preventive optimal strategies: the use of nose masks and vaccination. Both strategies among several measures have been adopted by several countries to limit the spread of the disease. As the world strives to get an effective medication to control the disease, appropriate control measures, for now, remains one of the effective measures to reduce the spread of the disease [54].

## 2 Model Formulation

We examine the transmission dynamics of the COVID-19 in Ghana by modifying the model of [34] by introducing a compartment for vaccination and extending the period of simulation from March 2020 to March 2021. In the new model, we assumed the infectious class is either asymptomatic or symptomatic. We first, consider a non - optimal control model, examine the basic properties of the model and then compute the basic reproduction number and later incorporate the optimal controls in the formulated model. We partition the population into six (6) compartments, namely: Susceptible individuals (S), Exposed (E), Asymptomatic (A), Symptomatic (Q), Vaccinated (V) and Recovered (R).

We assumed that the population is homogeneously mixed with no restriction on age and other social factors. Once infected, you become exposed to the disease before becoming infectious. For the model, only the asymptomatic individuals transmit the virus when they come in contact with the susceptible. Individuals are recruited into the susceptible class at the rate $\Omega$ and they die at the rate $\mu$. The transmission rate is $\beta$ and the disease-induced death rate is $\delta$. The parameters $\rho$ and $\gamma$ are the recovery rate of asymptomatic and symptomatic individuals respectively. The flowchart of the model is shown in Fig. 1.

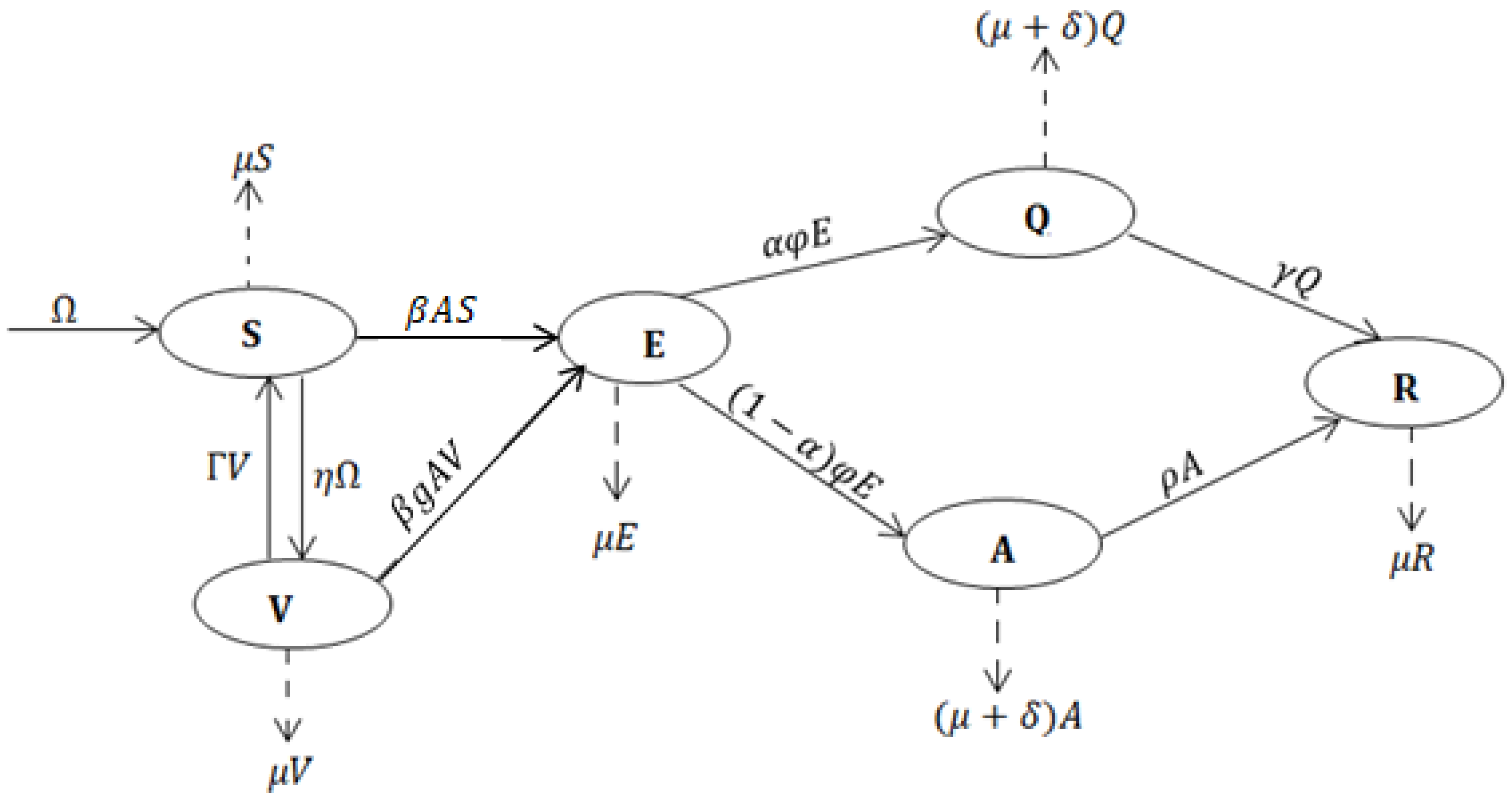

Fig. 1: Flowchart of the COVID-19 model

The following ordinary differential equations describe the model:

$$\begin{cases} \dfrac{dS}{dt} = (1-\eta)\Omega + \Gamma V - \dfrac{\beta AS}{N} - \mu S, \\ \dfrac{dE}{dt} = \dfrac{\beta AS}{N} + \dfrac{\beta gAV}{N} - (\varphi+\mu)E, \\ \dfrac{dA}{dt} = \alpha\varphi E - (\rho+\mu+\delta)A, \\ \dfrac{dQ}{dt} = (1-\alpha)\varphi E - (\gamma+\mu+\delta)Q, \\ \dfrac{dV}{dt} = \eta\Omega - \dfrac{\beta gAV}{N} - (\Gamma+\mu)V, \\ \dfrac{dR}{dt} = \rho A + \gamma Q - \mu R. \end{cases} \tag{1}$$

With initial conditions $S(0)=S_0, E(0)=E_0, A(0)=A_0, Q(0)=Q_0, V(0)=V_0, R(0)=R_0$.

We present the following results which guarantee that system (1) is epidemiologically and mathematically well-posed in a feasible region $\psi$ [51], given as

$$\psi = \left[ (S, E, A, Q, V, R) \in R_+^6 : N \leq \frac{\Omega}{\mu} \right]. \qquad (2)$$

**Theorem 1**: There exists a domain $\psi$ in which the solution set $(S, E, A, Q, V, R)$ is contained and bounded [51]

Proof:

Given the solution set $(S, E, A, Q, V, R)$ with positive initial conditions

$$S(0) = S_o, E(0) = E_o, A(0) = A_o, Q(0) = Q_o, V(0) = V_o, R(0) = R_o,$$

We let , $N(t) = S(t) + E(t) + A(t) + Q(t) + V(t) + R(t),\ then$

$$N'(t) = S'(t) + E'(t) + A'(t) + Q'(t) + V'(t) + R'(t).$$

It follows that $N'(t) < \Omega - \mu N.$

Solving the differential inequalities yields

$$N'(t) \leq \frac{\Omega}{\mu} + N(0)e^{-\mu(t)}.$$

Taking the limits as $t \to \infty$, gives $N' \leq \frac{\Omega}{\mu}.$

That is, all solutions are confined in the feasible region $\psi$ . We now show that the solutions of system (1) are non-negative in $\psi$ .

**Theorem 2:** If the initial states are non-negative, then they remain so for all $t > 0$ in the domain $\psi$ [51].

*Proof:* Clearly it is easy to see $S(t) > 0$, for all $t \geq 0$. If not, let there exist $t_* > 0$ such that $S(t_*) = 0, S'(t_*) \leq 0$ for all $0 \leq t \leq t_*$. Then from the first equation of system (1), we have $S'(t_*) = (1-\eta)\Omega \geq 0$, which is a contradiction, hence $S(t) \geq 0$

Moreover, from the second equation in system (1), it is easy to see $E(t) \geq 0$ for all $t \geq 0$. If not let there exist $t_* \geq 0$ such that $E(t_*) = 0$ and $E'(t_*) \leq 0$ for all $0 \leq t \leq 0$ which is a contradiction, hence, $E'(t) \geq 0$ . The same can be said about $V(t) \geq 0$. Furthermore from the third equation in system (1), it is easy to see $A(t) \geq 0$ for all $t \geq 0$. If not let there exist $t_* \geq 0$ such that $A(t_*) = 0$ *and* $A'(t_*) \leq 0$ for all $0 \leq t \leq 0$ we have $A'(t_*) = 0$ which is a contradiction, hence $A(t) \geq 0$. For Q(t), suppose $t_* \geq 0$, there exist $Q(t_*) = 0$ and $Q(t) > 0$, where $t \in [0, t_*)$, then from the fourth equation in system (1) we have

$$Q(t_*) = Q(0)e^{-(\gamma+\mu+\delta)t} + e^{(\gamma+\mu+\delta)t}\int_0^{t_*}(1-\alpha)\varphi E(\tau)e^{(\gamma+\mu+\delta)\tau}d\tau > 0 \text{ which contradicts } Q(t_*) = 0$$

Similarly, for R(t) assume that there exist $t_* \geq 0$ such that $R(t_*) = 0$ and $R(t) \geq 0$. Then solving the sixth equation of system (1) we obtain $R(t_*) = R(0)e^{-\mu(t_*)} + e^{-\mu(t_*)}\int_0^{t_*}[\rho A(\tau) + \gamma Q(\tau)]e^{\mu(\tau)}d\tau > 0$ , which contradicts $R(t_*) = 0$. Hence, this completes the proof.

## 3 Analysis of the Model

The disease – free equilibrium $(E_0)$ is the steady state solution where there is no infection in the population. Thus

$$E^0 = (S^0, E^0, A^0, Q^0, V^0, R^0) = \left( \frac{\Omega(\Gamma + \mu(1-\eta))}{\mu(\mu+\Gamma)}, 0, 0, 0, \frac{\eta\Omega}{(\Gamma+\mu)}, 0 \right) \qquad (3)$$

We now calculate the basic reproduction number $(R_0)$ of the model (1) using the next generation operator method [19]. Denoting F and V, respectively, as matrices for the new infections generated and the transition terms, thus

$$F = \begin{pmatrix} 0 & \frac{\beta\Omega(\Gamma+\mu(1-\eta))}{\mu(\mu+\Gamma)} & 0 \\ \alpha\varphi & 0 & 0 \\ (1-\alpha)\varphi & 0 & 0 \end{pmatrix}, \quad V = \begin{pmatrix} \varphi+\mu & 0 & 0 \\ 0 & \rho+\mu+\delta & 0 \\ 0 & 0 & \gamma+\mu+\delta \end{pmatrix}.$$

$$V^{-1} = \frac{1}{(\varphi+\mu)(\rho+\mu+\delta)(\gamma+\mu+\delta)} \begin{pmatrix} (\rho+\mu+\delta)(\gamma+\mu+\delta) & 0 & 0 \\ 0 & (\varphi+\mu)(\gamma+\mu+\delta) & 0 \\ 0 & 0 & (\varphi+\mu)(\rho+\mu+\delta) \end{pmatrix},$$

$$V^{-1} = \begin{pmatrix} \frac{1}{(\varphi+\mu)} & 0 & 0 \\ 0 & \frac{1}{(\rho+\mu+\delta)} & 0 \\ 0 & 0 & \frac{1}{(\gamma+\mu+\delta)} \end{pmatrix}.$$

$$FV^{-1} = \begin{pmatrix} 0 & 0 & 0 \\ \frac{\alpha\varphi}{\varphi+\mu} & \frac{\beta\Omega(\Gamma+\mu(1-\eta))}{\mu(\mu+\Gamma)(\rho+\mu+\delta)} & 0 \\ \frac{(1-\alpha)\varphi}{\varphi+\mu} & 0 & 0 \end{pmatrix}.$$

The basic reproduction number is the largest positive eigenvalue of $FV^{-1}$, i.e.

$$R_0 = \frac{\beta\Omega(\Gamma + \mu(1-\eta))}{\mu(\mu+\Gamma)(\rho+\mu+\delta)}. \qquad (4)$$

To calculate $R_0$ at the early stage of the infection, we exclude the vaccination compartment and using the next generation operator method [19], $R_0$ is given as

$$R_0 = \frac{\beta}{(\rho+\mu+\delta)}. \qquad (5)$$

The necessary condition for the local stability of the disease –free steady state is established in the following theorem.

**Theorem 3:** The disease-free equilibrium is locally asymptotically stable if $R_o < 1$ and unstable for $R_o > 1$ [51].

**Proof:**

The Jacobian matrix of system (2) is given as

$$J = \begin{bmatrix} -\mu-\beta A & 0 & -\beta S & 0 & \Gamma & 0 \\ \beta A & -(\varphi+\mu) & \beta S+\beta g V & 0 & \beta g A & 0 \\ 0 & \alpha\varphi & -(\rho+\mu+\delta) & 0 & 0 & 0 \\ 0 & (1-\alpha)\varphi & 0 & -(\gamma+\mu+\delta) & 0 & 0 \\ 0 & 0 & -\beta g V & 0 & -\beta g A-(\Gamma+\mu) & 0 \\ 0 & 0 & \rho & \gamma & 0 & -\mu \end{bmatrix}. \qquad (6)$$

The Jacobian matrix evaluated at the disease-free equilibrium point is

$$J_{E^O} = \begin{bmatrix} -\mu & 0 & -\beta S^o & 0 & \Gamma & 0 \\ 0 & -(\varphi+\mu) & \beta S^o & 0 & 0 & 0 \\ 0 & \alpha\varphi & -(\rho+\mu+\delta) & 0 & 0 & 0 \\ 0 & (1-\alpha)\varphi & 0 & -(\gamma+\mu+\delta) & 0 & 0 \\ 0 & 0 & -\beta g V^0 & 0 & -(\Gamma+\mu) & 0 \\ 0 & 0 & \rho & \gamma & 0 & -\mu \end{bmatrix}. \qquad (7)$$

We show that all eigenvalues of system (7) are negative. The first, fourth, fifth, and sixth columns give the first four (4) eigenvalues which are $-(\gamma+\mu+\delta)$, $-(\Gamma+\mu)$ and $-\mu$ (repeated roots). The rest are obtained from the $(2\times 2)$ sub-matrix formed by excluding the first, fourth, fifth and sixth rows and columns of system (7). Hence we have

$$J_{E^0} = \begin{pmatrix} -(\varphi+\mu) & \beta S^0 \\ (1-\alpha)\varphi & -(\rho+\mu+\delta) \end{pmatrix}. \qquad (8)$$

The characteristic equation of system (8) is given as $\lambda^2 + a_1\lambda + a_2 = 0$,

where

$$a_1 = (\varphi+\rho+2\mu+\delta),$$
$$a_2 = (\varphi+\mu)(\rho+\mu+\delta) - \frac{(1-\alpha)\varphi}{(\rho+\mu+\delta)} R_0.$$

From Routh – Hurwitz stability criterion if the condition, $a_1 > 0 \; and \; a_2 > 0$ are satisfied then the equilibrium point is stable. The coefficients of the characteristic equation given by $a_1$ and $a_2$ are greater than zero for $R_0 < 1$ which then makes the disease-free equilibrium point stable. For $R_0 > 1$, $a_2 < 0$ and this makes the disease-free equilibrium point unstable.

### 3.1 Existence of the Endemic Equilibrium Point and Local Stability

We denote the endemic equilibrium point by $E_1 = (S^*, E^*, V^*, A^*, Q^*, R^*)$ , equating the right hand – side of the system (1) to zero and solving yields

$$S^* = \frac{(1-\eta)\Omega + \Gamma V^*}{\beta A^* + \mu}, E^* = \frac{\beta S^* A^* + \beta g A^* V^*}{\varphi+\mu}, A^* = \frac{\alpha\varphi E^*}{\rho+\mu+\delta},$$
$$Q^* = \frac{(1-\alpha)\varphi E^*}{\gamma+\mu+\delta}, V^* = \frac{\eta\Omega}{\beta g A^* + (\Gamma+\mu)}, R^* = \frac{\rho A^* + \gamma Q^*}{\mu}. \qquad (9)$$

We state and prove the following theorem:

**Theorem 4:** If $R_0 > 1$, then the endemic equilibrium point $E^*$, is locally asymptotically stable [51].

**Proof:**

The Jacobian matrix (6) evaluated at the endemic equilibrium point is given as

$$J_{E^*} = \begin{bmatrix} -\mu-\beta A^* & 0 & -\beta S^* & 0 & \Gamma & 0 \\ \beta A^* & -(\varphi+\mu) & \beta S^* + \beta g V^* & 0 & \beta g Q^* & 0 \\ 0 & \alpha\varphi & -(\rho+\mu+\delta) & 0 & 0 & 0 \\ 0 & (1-\alpha)\varphi & 0 & -(\gamma+\mu+\delta) & 0 & 0 \\ 0 & 0 & -\beta g V^* & 0 & -\beta g A^* - (\Gamma+\mu) & 0 \\ 0 & 0 & \rho & \gamma & 0 & -\mu \end{bmatrix}. \quad (10)$$

The first two eigenvalues can be obtained from the fourth and the sixth columns which are $-(\gamma+\mu+\delta)$ and $-\mu$ . The remaining eigenvalues are obtained by excluding the fourth and sixth rows and columns from system (10) and this gives

$$J_{E^*} = \begin{bmatrix} -(\mu+\beta A^*) & 0 & -\beta S & \Gamma \\ \beta A^* & -(\varphi+\mu) & \beta S^* + \beta g V^* & \beta g A^* \\ 0 & (1-\alpha)\varphi & -(\gamma+\mu+\delta) & 0 \\ 0 & 0 & -\beta g V^* & -\beta g A^* - (\Gamma+\mu) \end{bmatrix}. \quad (11)$$

The characteristic equation of system (11) is

$$\lambda^4 + A_1\lambda^3 + A_2\lambda^2 + A_3\lambda + A_4 = 0, \quad (12)$$

where

$$A_1 = (J_{11} + J_{22} + J_{33} - J_{44}),$$
$$A_2 = (J_{11}J_{22} + J_{11}J_{33} + J_{22}J_{33} - J_{32}J_{23} - J_{11}J_{44} - J_{22}J_{44} - J_{33}J_{44}),$$
$$A_3 = J_{11}(J_{22}J_{33} - J_{32}J_{23} - J_{22}J_{44} - J_{33}J_{44}) - J_{44}(J_{22}J_{33} - J_{23}J_{32}) - J_{32}\beta^2 g^2 Q^*,$$
$$A_4 = J_{32}J_{44}(J_{11}J_{23} + \beta^2 S^* A^*) - J_{32}\beta g(\Gamma\beta A^* + J_{11}\beta g A^*) - J_{11}J_{22}J_{33}J_{44},$$
$$J_{11} = \mu + \beta A^*, J_{13} = \beta S^*, J_{22} = \varphi + \mu, J_{32} = (1-\alpha)\varphi, J_{23} = \beta(S^* + gV^*),$$
$$j_{33} = \gamma + \mu + \delta, J_{44} = -\beta g A^* - (\Gamma + \mu).$$

From Routh – Hurwitz stability criterion, if the conditions $A_1 > 0, A_3 > 0 \quad A_4 > 0$ and $A_1A_2A_3 > A_3^2 + A_1^2A_4$ are satisfied, then the characteristic equation above has negative real parts and hence a stable equilibrium.

### 3.2 Global stability of the disease-free equilibrium

We state the following theorem with proof.

**Theorem 5:** The disease-free equilibrium of the system (1) is globally asymptotically stable if $R_o < 1$ and unstable otherwise.

**Proof:** Applying the theorem on [20], we separate the system (1) into infectious and non-infectious classes denoted by $M = (E, A, Q) \in R_+^3$ and $N = (S, V, R) \in R_+^3$ respectively. The system (1) can now be written as

$$\frac{dM}{dt} = V(M,N), \frac{dN}{dt} = G(M,N) \qquad (13)$$

The two valued function $V(M,N$ and $G(M,N)$ are given by

$$V(M,N) = \begin{pmatrix} \beta AS + \beta gAV - (\varphi + \mu)E \\ \alpha\varphi E - (\rho + \mu + \delta)A \\ (1-\alpha)\varphi E - (\gamma + \mu + \delta)Q \end{pmatrix}, \quad G(M,N) = \begin{pmatrix} (1-\eta)\Omega + \Gamma V - \beta AS - \mu S \\ \eta\Omega - \beta gAV - (\Gamma + \mu)V \\ \rho A + \gamma Q - \mu R \end{pmatrix}$$

The reduced form of the system $dN/dt = G(N,0)$ is given as

$$\frac{dS}{dt} = (1-\mu)\Omega + \Gamma V - \mu S,$$
$$\frac{dV}{dt} = \eta\Omega - (\Gamma + \mu)V, \qquad (14)$$
$$\frac{dR}{dt} = -\mu R,$$

$$N^* = (S^*, V^*, R^*) = \left( \frac{(1-\eta)\Omega + \Gamma V^*}{\mu}, \frac{\eta\Omega}{\Gamma + \mu}, 0 \right),$$

which is globally asymptotically stable equilibrium point for the reduced system

$dN/dt = G(N,0)$. The third equation of system (14) gives $R(t) = R(0)e^{-\mu(t)} \to 0$ as $t \to \infty$.

Solving the second equation of system (14) gives $V(t) = \frac{\eta\Omega}{\Gamma+\mu} + V(0)e^{-(\Gamma+\mu)t} \to \frac{\eta\Omega}{\Gamma+\mu}$ as $t \to \infty$

. Solving the first equation of system (14) gives

$$S(t) = \frac{1}{\mu}\left[ (1-\mu)\Omega + \Gamma\left( \frac{\eta\Omega}{\Gamma+\mu} + V(0)e^{-(\Gamma+\mu)t} \right) \right] + S(0)e^{-\mu(t)} \to \frac{1}{\mu}\left[ (1-\mu)\Omega + \Gamma\left( \frac{\eta\Omega}{\Gamma+\mu} \right) \right] \text{ as}$$

$t \to \infty$

Hence, the convergence of system (2) is global in $\psi$ $.V = (M, N)$, satisfies the following conditions given in [20], i.e.

1. $\frac{dM}{dt} = V(M,0) = 0,$

2. $V = (M, N) = AM - \mathrm{T}(M, N) \geq 0,$

where $A = \Phi_M V(M,0) = \begin{pmatrix} \beta g A V - (\varphi + \mu)E \\ \alpha \varphi E - (\rho + \mu + \delta)A \\ (1-\alpha)\varphi E - (\gamma + \mu + \delta)Q \end{pmatrix}$ satisfies the conditions above.

## 4 Numerical Analysis of the COVID-19 Model

After formulating a model, one important thing is to validate the model to see if it will stand the test of time. Model validation is the process of determining the degree to which a mathematical model is an accurate representation of the available data. In this section, we validate the model by using cumulative cases of Ghana from Ghana Health Service for the period March 2020 to March, 2021 [24]. We also estimate the parameters of the model and test the effect of the model parameters on the basic reproductive number. The cumulative data of confirmed COVID-19 cases for the period March 2020 to March, 2021 is depicted in figure 2 and figure 3 shows the residuals of the best fitted curve.

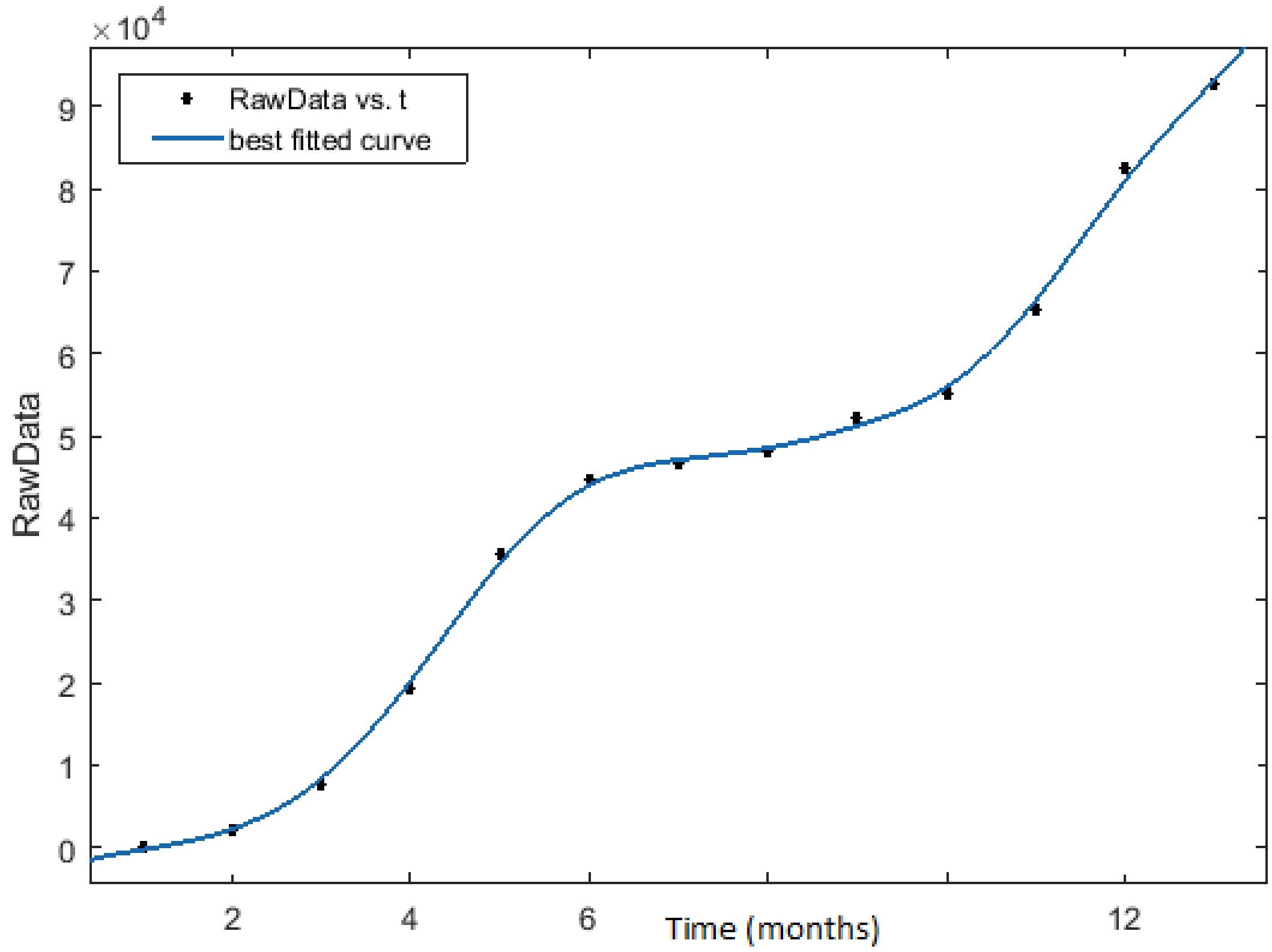

Fig.2: Cumulative cases of Ghana's COVID-19 from March to June 2020 with the best fitted curve.

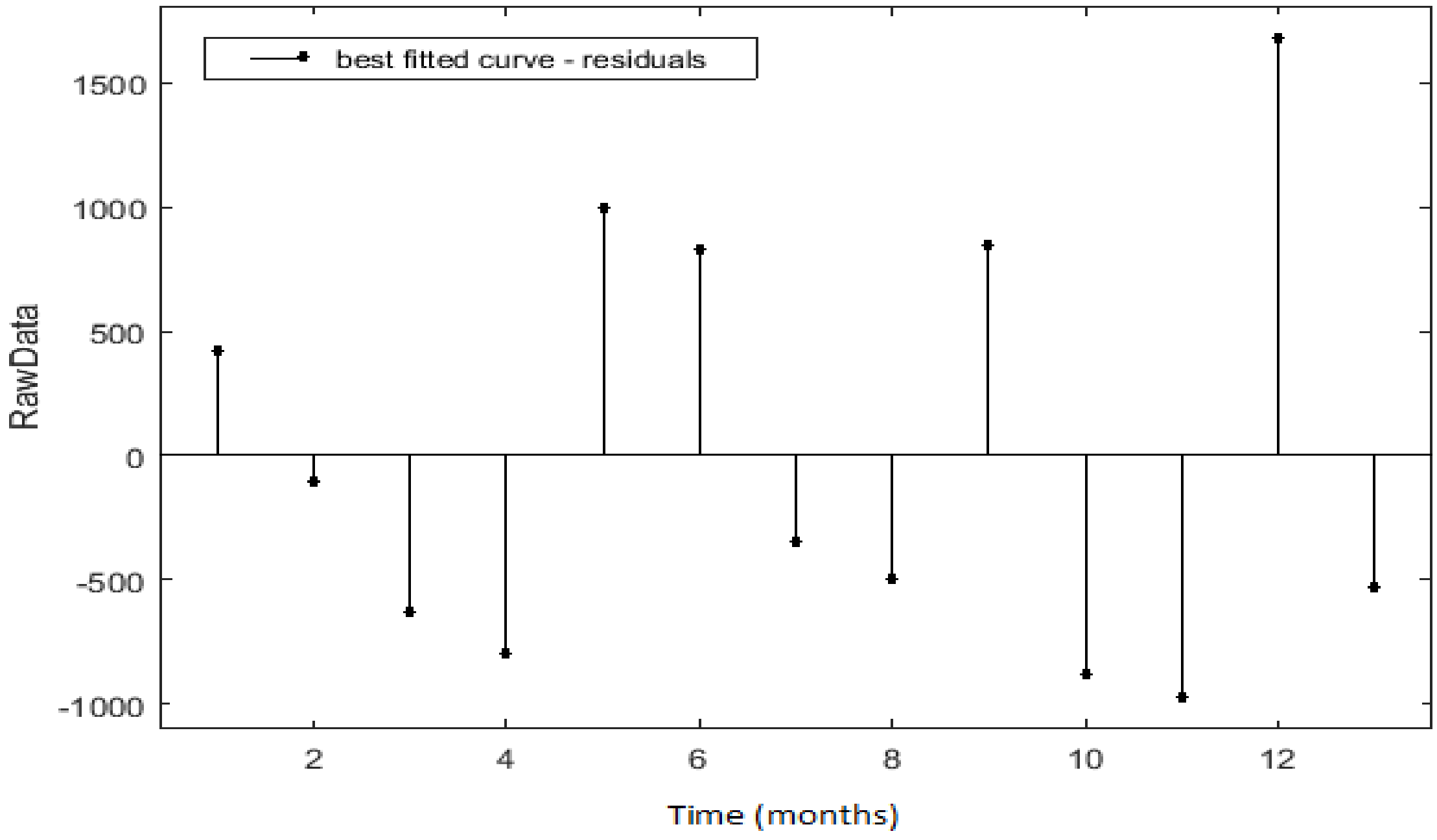

Fig.3: Residuals for the best fitted curve

We now introduce another compartment (D) into the model to cater for deaths due to COVID-19 disease. Hence system (2) becomes

$$\begin{aligned}
\frac{dS}{dt} &= (1-\eta)\Omega + \Gamma V - \frac{\beta AS}{N} - \mu S, \\
\frac{dE}{dt} &= \frac{\beta AS}{N} + \frac{\beta gAV}{N} - (\varphi + \mu)E, \\
\frac{dA}{dt} &= \alpha\varphi E - (\rho + \mu + \delta)A, \\
\frac{dQ}{dt} &= (1-\alpha)\varphi E - (\gamma + \mu + \delta)Q, \\
\frac{dV}{dt} &= \eta\Omega - \frac{\beta gAV}{N} - (\Gamma + \mu)V, \\
\frac{dR}{dt} &= \rho A + \gamma Q - \mu R, \\
\frac{dD}{dt} &= \delta(A+Q)
\end{aligned} \tag{15}$$

A starting point of our simulation is March 2020 where the authorities of Ghana confirmed the first two cases of the COVID-19 [25]. According to 2020 population and housing census, the population of Ghana is 30.8 million [53]. We choose these initial conditions $S(0) = N - A(0), E(0) = 0, A(0) = 2, Q(0) = 0, V(0) = 0, R(0) = 0$.

The natural death rate $\mu$ is equal to the inverse of the life expectancy at birth. The country's life expectancy for 2021 is 64.42 [51], so $\mu = \frac{1}{(64.42)(365)} = 0.4252912 \times 10^{-4}$ per day. Usually, it takes 14 days for a person with mild symptoms to recover from COVID-19 through a proper treatment process. The current birth rate for Ghana in 2021 is 29.08 births per 1000 people [21]. Persons fully vaccinated as of 31st March 2021 was 500,000 [36]. The parameter values are given in Table 1.

Table 1: Parameter values and description

| Parameter | Description | Value | Reference |
|---|---|---|---|
| $\eta$ | Rate at which susceptible individuals are vaccinated | 0.01624 | Estimated |
| $\Omega$ | Recruitment rate | 29.08 | [21] |
| $\mu$ | Natural death rate | $0.4252912\times10^{-4}$ | Estimated |
| $\delta$ | Disease-induced death rate | $1.6728\times10^{-5}$ | [22] |
| $\varphi$ | The rate at which exposed individuals become infectious | 0.25 | [35] |
| $\rho$ | The recovery rate of asymptomatic individuals | 1/14 | Estimated |
| $\Gamma$ | Vaccine waning rate | 0.0000001 | Assumed |
| $\beta$ | Probability of human getting infected with covid-19 | 0.9 | Estimated |
| $\gamma$ | The recovery rate of symptomatic individuals | 1/14 | Estimated |

We now determine which of the parameters are most influential to control the basic reproductive number.

**4.1 Sensitivity Analysis of COVID-19 Model**

We perform a sensitivity analysis on the parameters of system (2) to determine which parameter will increase or decrease the basic reproductive number. We use the normalized forward sensitivity index given as $X = \frac{\partial R_o}{\partial \theta} \times \frac{\theta}{R_o}$, where $\theta$ is the parameter under consideration. Positive sensitivity index means an increase in that parameter will lead to an increase in the basic reproductive number and a negative sensitivity index means an increase in the parameter will decrease the basic reproductive number. We use the parameter values given in Table 1 to determine the sensitivity indices in Table 2 below.

Table 2: Sensitivity Indices of the model parameters

| Parameter | Value | Sensitivity Indices |
|---|---|---|
| $\mu$ | $0.4252912\times10^{-4}$ | -1.7008e-04 |
| $\delta$ | $1.6728\times10^{-5}$ | -6.6896e-05 |
| $\varphi$ | 0.017199 | 0.017199 |
| $\rho$ | 1/4 | -0.9998 |
| $\beta$ | 0.936862 | 3.5985 |

From Table 2, it can be seen that the most positive sensitivity indices is the parameter $\beta$. On the other hand, the most negative sensitivity index is the parameter $\rho$ which is the recovery rate for the asymptomatic individuals.

**4.2 Numerical Simulation**

Using the initial conditions S(0)=30,800000, E(0)=0, A(0)=02, Q(0)=0, V(0)=0, R(0)=0, the simulations performed are displayed in Figs. 4 – 10 which depicts the behaviour of all the compartment for the first 400 days since the outbreak, i.e., from March 2020 to March 2021.

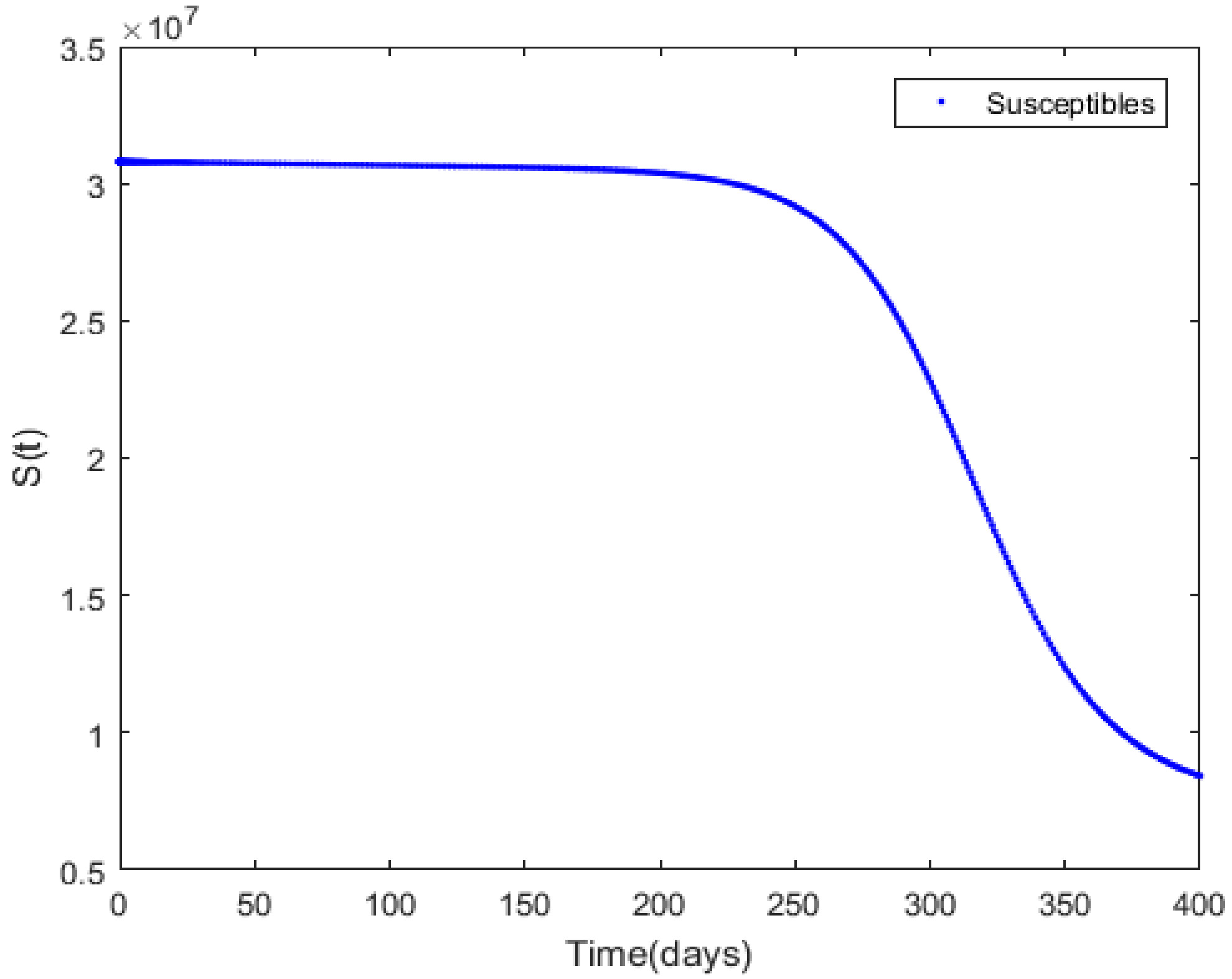

Fig. 4: Behaviour of susceptible individuals

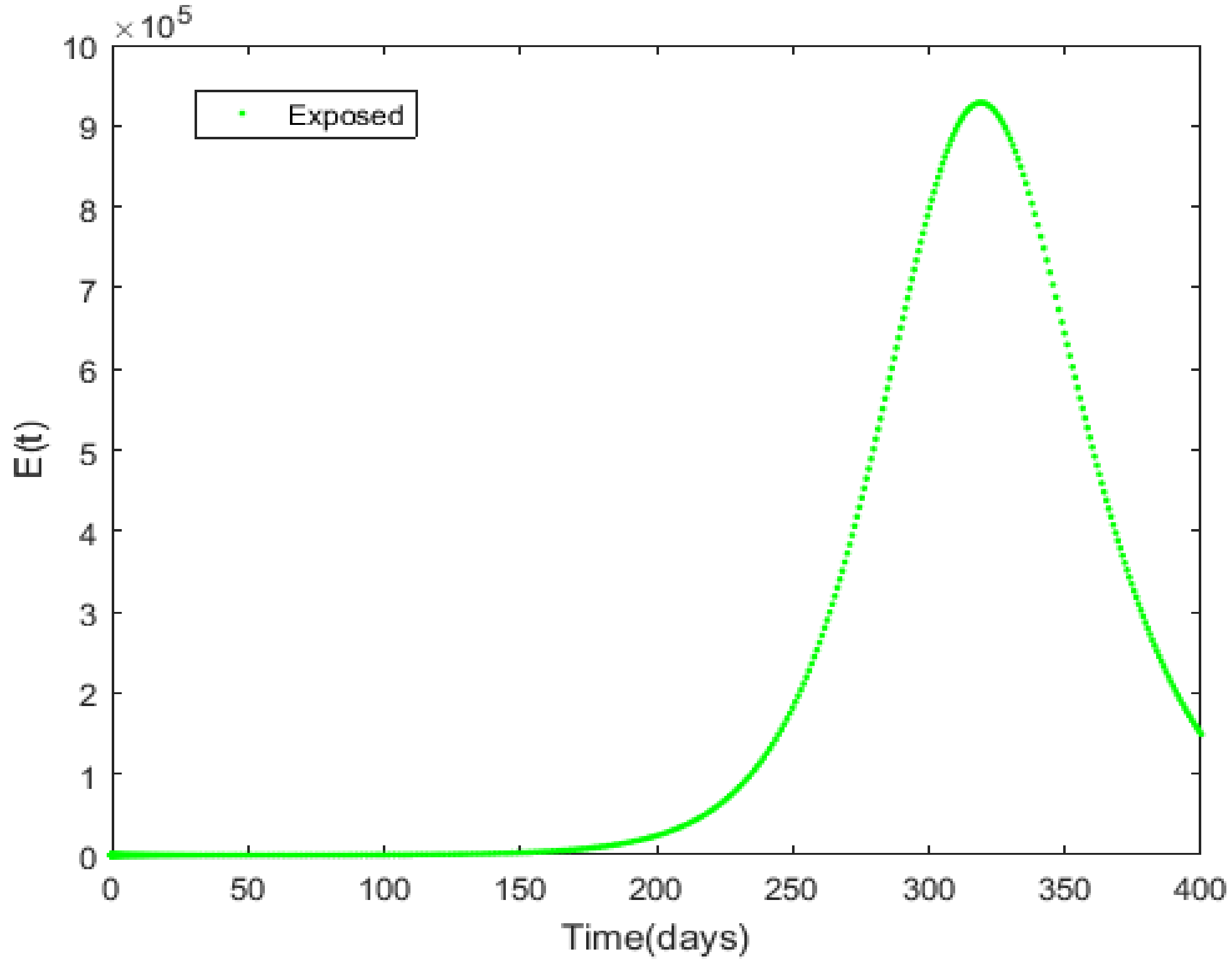

Fig. 5: Behaviour of the exposed individuals

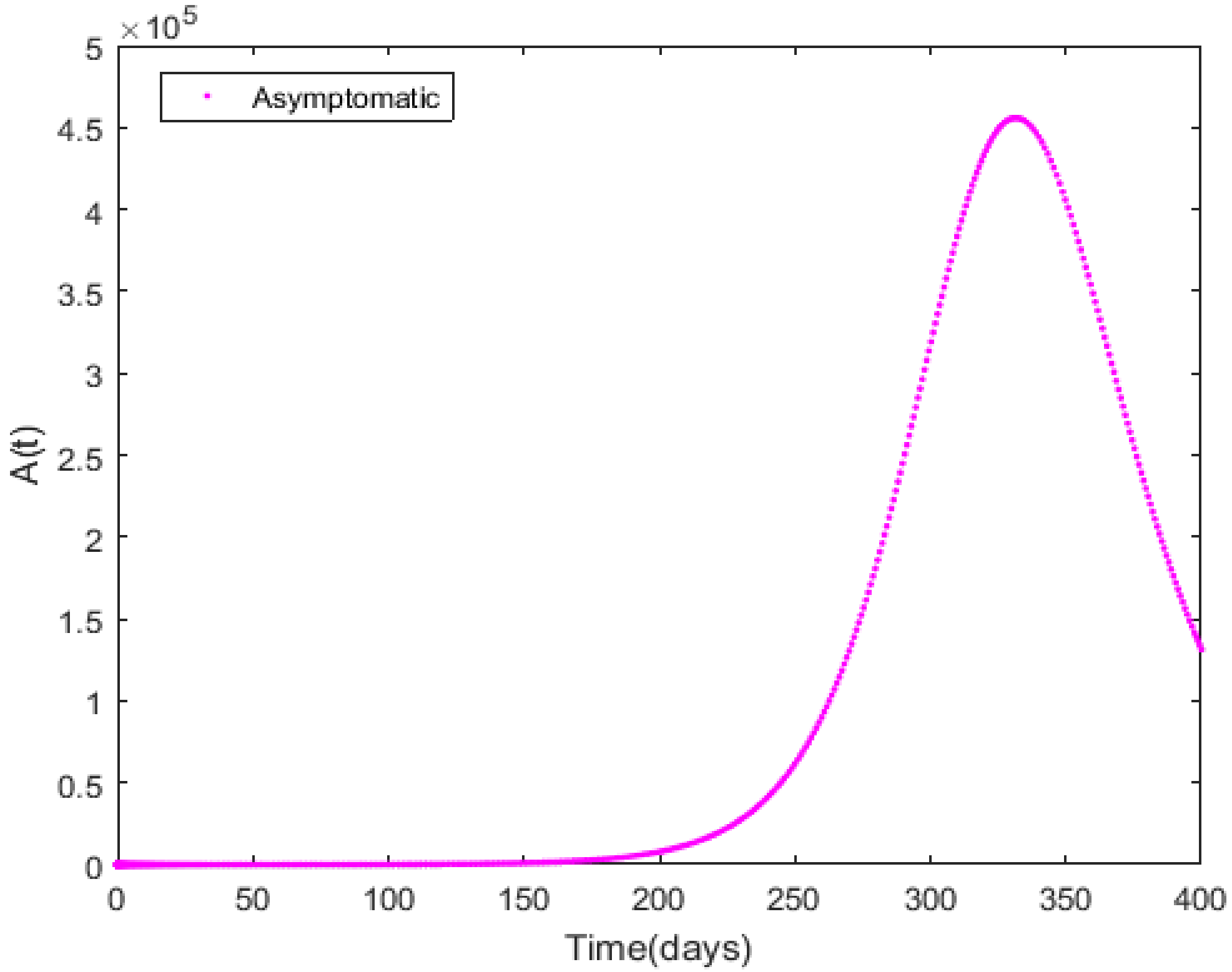

Fig.6: Behaviour of asymptomatic individuals

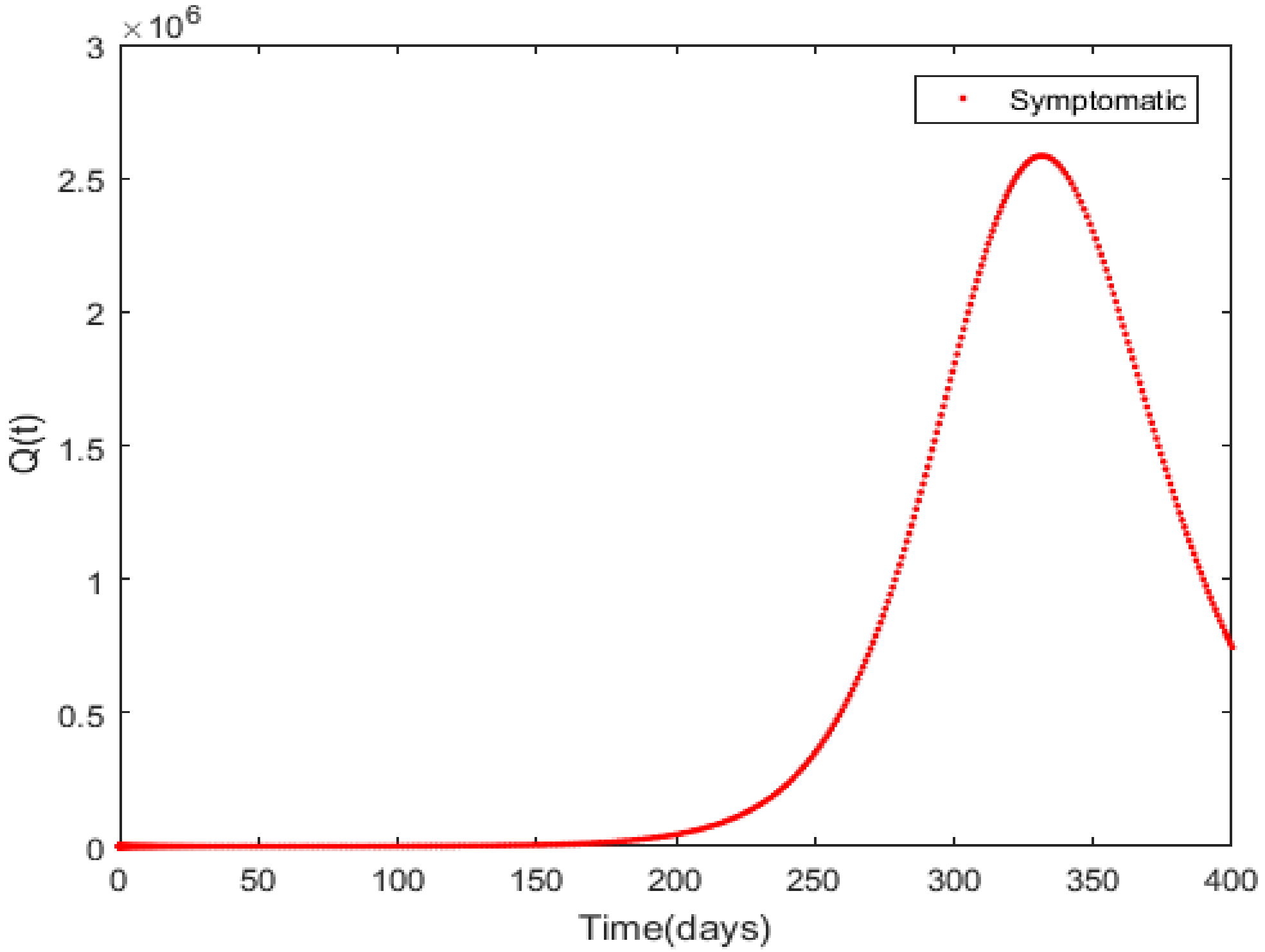

Fig.7: Behaviour of the symptomatic individuals

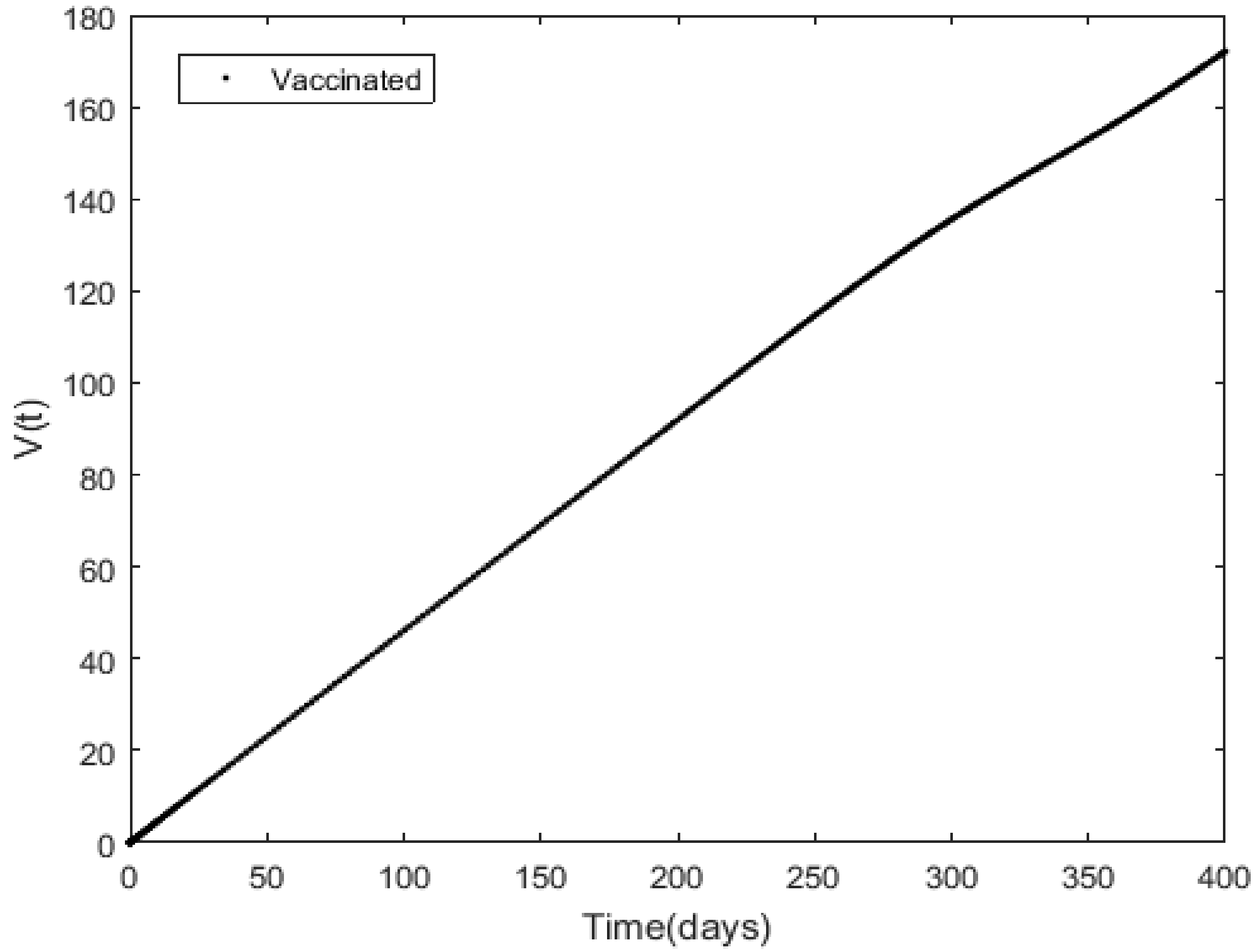

Fig. 8: Behaviour of the vaccinated class

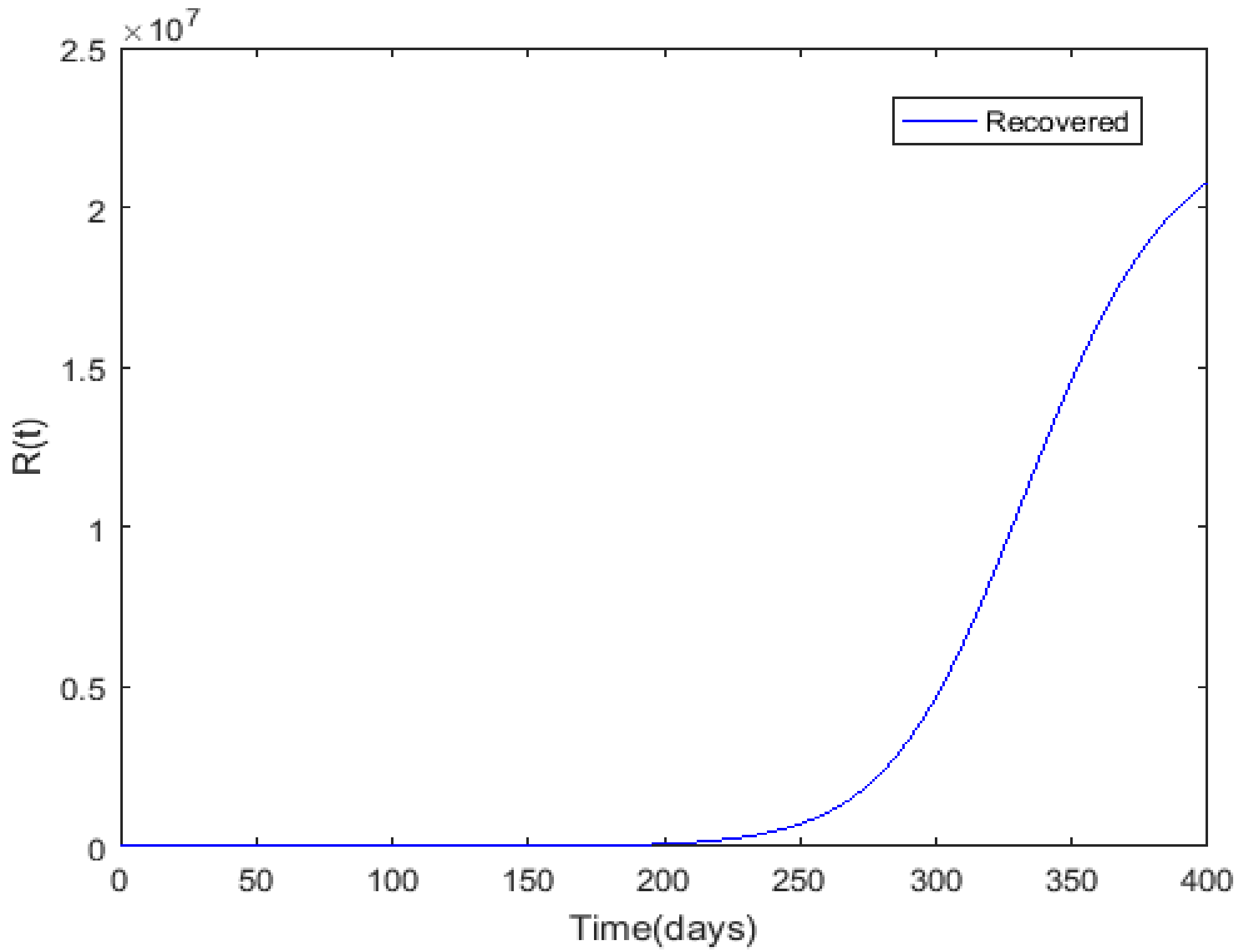

Fig. 9: Behaviour of the recovered individuals

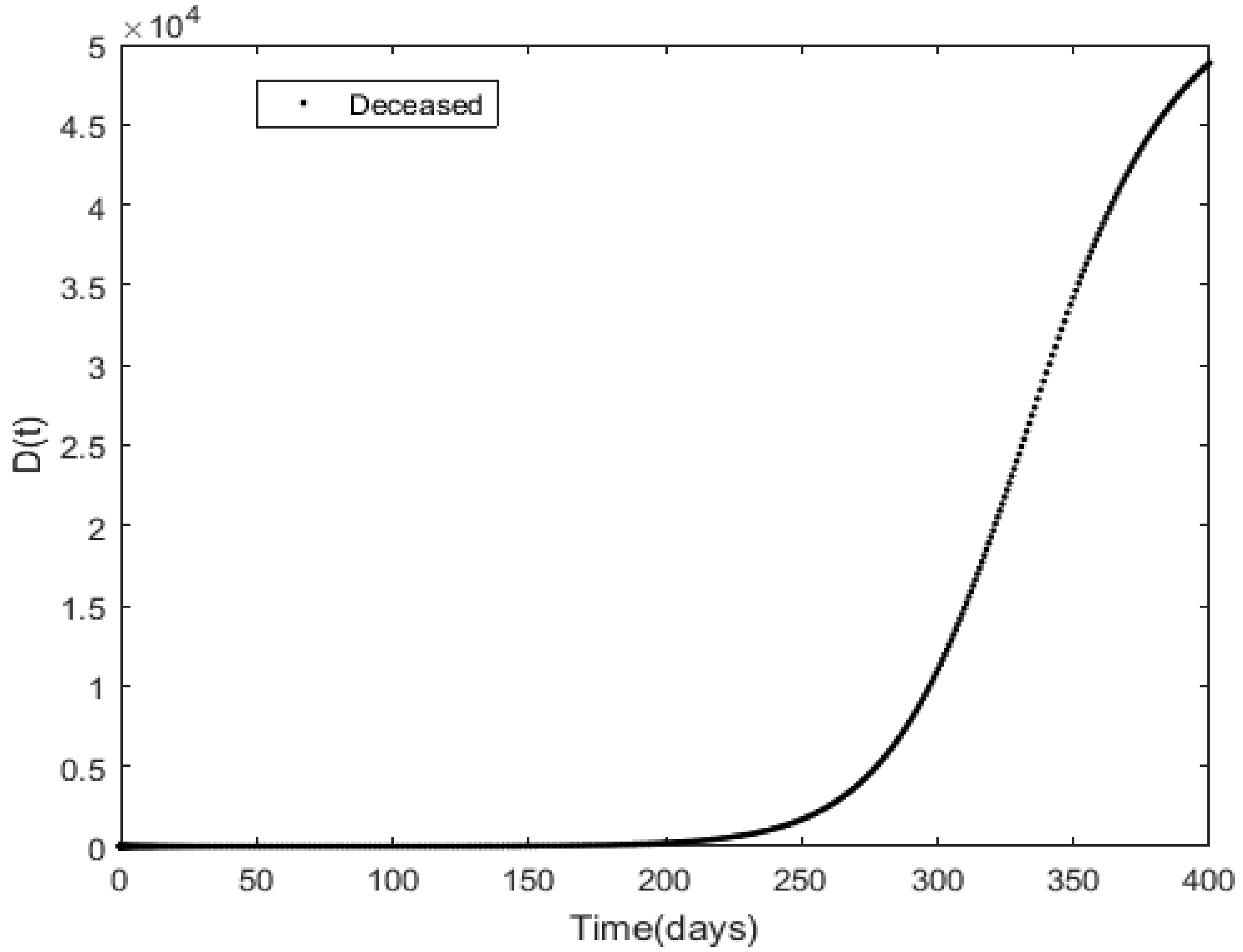

Fig.10: Behaviour of the deceased compartment

In Fig. 4, a decline in the population of the susceptible is noticed after the first 300 days of the outbreak. The number of exposed, asymptomatic and symptomatic individual increases after the first 200 days of the outbreak as shown in Figs. 5 – 7 respectively. In Fig. 8, the vaccinated individuals increase with time. There is a rapid increase in number of individuals who recovers from the disease and those deceased after the first 200 days as depicted in Fig. 9 and 10 respectively.

**5 The Optimal Control Method**

In this section, we look at two preventive control mechanisms i.e. control $(u_1)$ which represents the use of a mask and control $(u_2)$ vaccination of individuals. The directive to use mask started right from the beginning of the outbreak whereas vaccination of individuals started at the beginning of March 2021. To include the nose mask control in the model, we replaced the parameter $\beta$ with $(1-u_1)\beta$ where, $0 \leq u_1 \leq 1$ . If there are no usages of the mask, then $u_1 = 0$ and if the entire population uses the mask then $u_1 = 1$.We incorporate the time-dependent controls into the system (1) and we have

$$
\begin{aligned}
\frac{dS}{dt} &= (1-\eta)\Omega + \Gamma V - (1-u_1)\beta AS - \mu S - u_2 S \\
\frac{dE}{dt} &= (1-u_1)\beta AS + (1-u_1)\beta gAV - (\varphi+\mu)E \\
\frac{dA}{dt} &= \alpha\varphi E - (\rho+\mu+\delta)A \\
\frac{dQ}{dt} &= (1-\alpha)\varphi E - (\gamma+\mu+\delta)Q \\
\frac{dV}{dt} &= \eta\Omega - (1-u_1)\beta gAV - (\Gamma+\mu)V + u_2 S \\
\frac{dR}{dt} &= \rho A + \gamma Q - \mu R
\end{aligned}
\tag{16}
$$

### 5.1 Analysis of the Optimal Control Technique

We analyse the behaviour of system (16). The objective function for fixed time $t_f$ is

$$J(u_1,u_2) = \int_0^{t_f} [g_1 S(t) + g_2 E(t) + g_3 A(t) + g_4 Q(t) + \frac{1}{2}(k_1 u_1^2 + k_2 u_2^2)]dt \tag{17}$$

Where $g_1, g_2, g_3, g_4$ are the relative weights and $k_1$ and $k_2$ are the relative cost associated with the controls $u_1$ and $u_2$. The final time of the control is $t_f$. The aim of the control is to minimize the cost function.

$$J(u_1^*,u_2^*) = \min_{u_1,u_2 \in U} J(u_1,u_2), \tag{18}$$

subject to system (16), where $0 \le (u_1, u_2) \le 1$ and $t \in (0, t_f)$. In other to derive the necessary condition for the optimal control, Pontryagin maximum principle given in [23] was used. This principle converts system (16) - (18) into a problem of minimizing a Hamiltonian H, defined by

$$
\begin{aligned}
H &= g_1 S(t) + g_2 E(t) + g_3 A(t) + g_4 Q(t) + \frac{1}{2}(k_1 u_1^2 + k_2 u_2^2) \\
&+ \Lambda_S \{(1-\eta)\Omega + \Gamma V - (1-u_1)\beta AS - \mu S - u_2 S\} \\
&+ \Lambda_E \{(1-u_1)\beta AS + (1-u_1)\beta gAV - (\varphi + \mu)E\} \\
&+ \Lambda_A \{\alpha\varphi E - (\rho + \mu + \delta)A\} \\
&+ \Lambda_Q \{(1-\alpha)\varphi E - (\gamma + \mu + \delta)Q\} \\
&+ \Lambda_V \{\eta\Omega - (1-u_1)\beta gAV - (\Gamma + \mu)V + u_2 S\} \\
&+ \Lambda_R \{\rho A + \gamma Q - \mu R\},
\end{aligned}
\tag{19}
$$

where $\Lambda_S, \Lambda_E, \Lambda_A, \Lambda_Q, \Lambda_V$ *and* $\Lambda_R$ represents the co-state variables. The system of equations is derived by taking into account the correct partial derivatives of system (19) with respect to the associated state variables.

**Theorem 6**: Given optimal control $u_1^*, u_2^*$ and corresponding solution $S^*, E^*, A^*, Q^*, V^*, R^*$ of the corresponding state system (16) – (17) that minimizes $J(u_1, u_2)$ over U, there exist adjoint variables $\Lambda_S, \Lambda_E, \Lambda_A, \Lambda_Q, \Lambda_V, \Lambda_R$, satisfying

$$
-\frac{d\Lambda_i}{dt} = \frac{\partial H}{\partial i}, \tag{20}
$$

where $i = \Lambda_S, \Lambda_E, \Lambda_A, \Lambda_Q, \Lambda_V, \Lambda_R$, with the transversality conditions

$$
\Lambda_S(t_f) = \Lambda_E(t_f) = \Lambda_A(t_f) = \Lambda_Q(t_f) = \Lambda_V(t_f) = \Lambda_R(t_f) = 0
$$

***Proof:*** The differential equations characterized by the adjoint variables are obtained by considering the right-hand side derivatives of system (23) determined by the optimal control. The adjoint equations derived are given as

$$
\begin{aligned}
\frac{d\Lambda_S}{dt} &= -g_1 + \beta A(1-u_1)[\Lambda_S - \Lambda_E] + \mu\Lambda_S + u_2[\Lambda_S - \Lambda_V] \\
\frac{d\Lambda_E}{dt} &= -g_2 + (\varphi+\mu)\Lambda_E - \alpha\varphi[\Lambda_Q - \Lambda_A] - \varphi\Lambda_Q, \\
\frac{d\Lambda_A}{dt} &= -g_3 + (1-u_1)\beta S[\Lambda_S - \Lambda_E] + (1-u_1)\beta gV[\Lambda_V - \Lambda_E] + (\rho+\mu+\delta)\Lambda_A - \rho\Lambda_R, \\
\frac{d\Lambda_Q}{dt} &= -g_4 + (\gamma+\mu+\delta)\Lambda_Q - \gamma\Lambda_R, \\
\frac{d\Lambda_V}{dt} &= -\Gamma\Lambda_S + (1-u_1)\beta gA[\Lambda_V - \Lambda_E] + (\Gamma+\mu)\Lambda_V, \\
\frac{d\Lambda_R}{dt} &= \mu\Lambda_R
\end{aligned}
\tag{25}
$$

By obtaining the solution for $u_1^*$ and $u_2^*$ subject to the constraints, we have

$$
\begin{aligned}
0 &= \frac{\partial H}{\partial u_1} = -k_1u_1 + \beta AS[\Lambda_E - \Lambda_S] + \beta gAV[\Lambda_E - \Lambda_V] \\
0 &= \frac{\partial H}{\partial u_1} = -k_2u_2 + S[\Lambda_S - \Lambda_V]
\end{aligned}
\tag{26}
$$

This gives

$$
\begin{aligned}
u_1^* &= \min\left(1, \max\left(0, \frac{\beta AS[\Lambda_E - \Lambda_S] + \beta gAV[\Lambda_E - \Lambda_V]}{k_1}\right)\right) \\
u_2^* &= \min\left(1, \max\left(\frac{S[\Lambda_S - \Lambda_V]}{k_2}\right)\right)
\end{aligned}
\tag{27}
$$

### 5.2 Numerical Analysis of the Optimal Control Model – Prevention Controls

In this section, we analysed numerically the behaviour of the optimal control model (16) using the method of forward-backward sweep method as in [45]. We develop a numerical scheme that uses matlab fourth order Runge-Kutta method [37 – 39, 45] to solve the model's optimality system. The control $u_1$ is optimized for the period whiles the control $u_2$ is set to zero. The results of the simulation are displayed in Figs. 11 – 16.

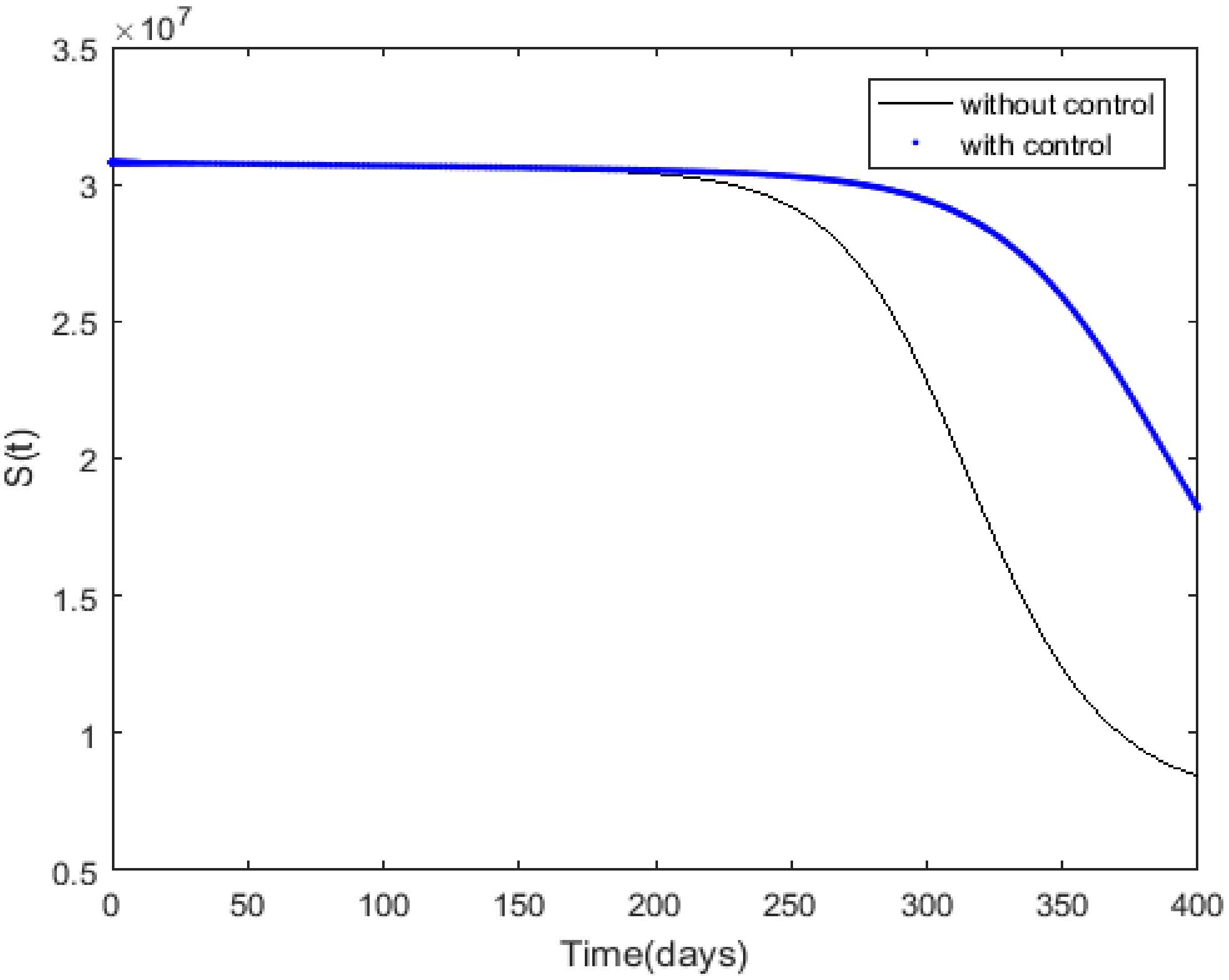

Fig. 11: Behaviour of the susceptible with and without control

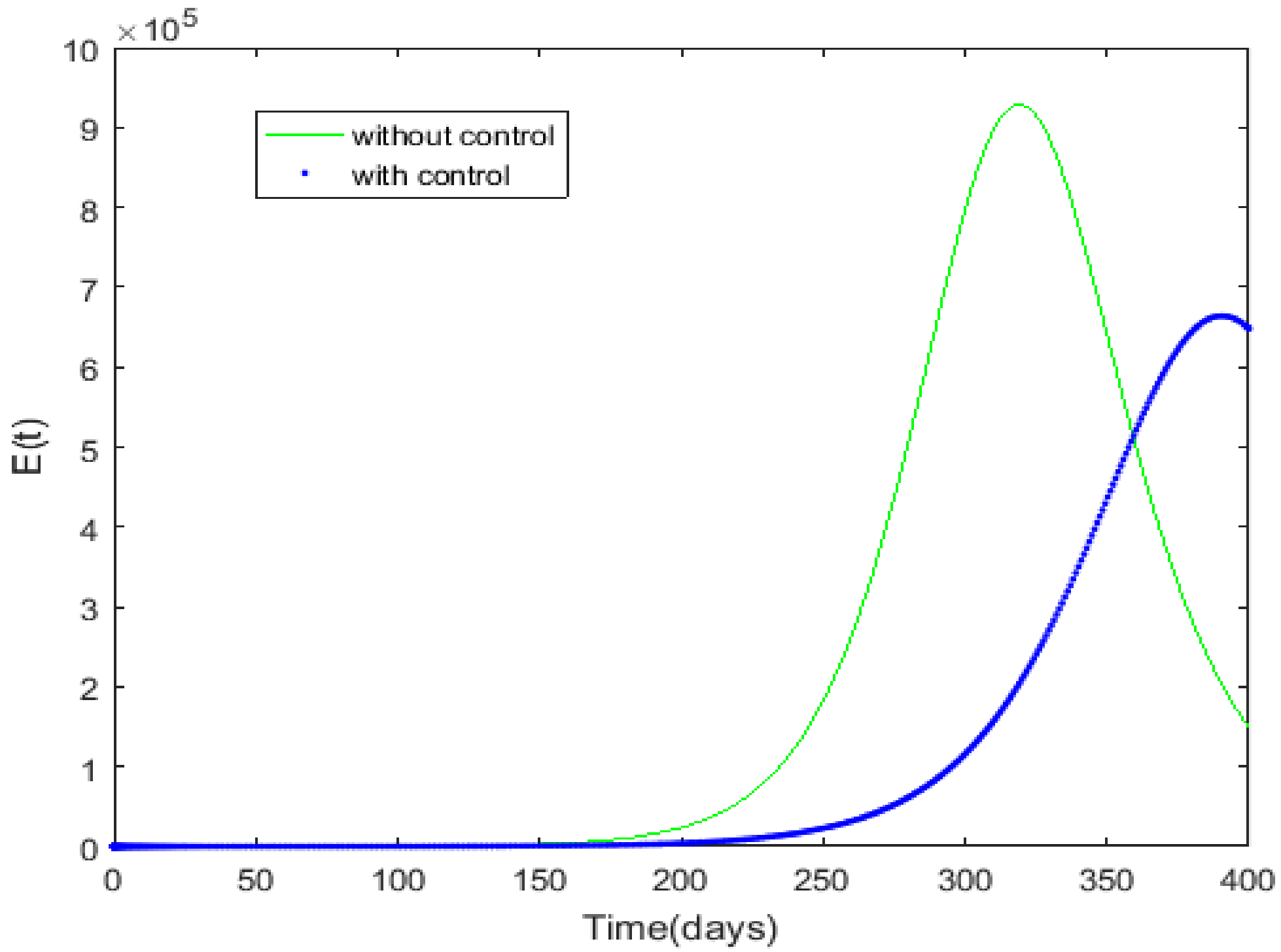

Fig. 12: Behaviour of the exposed individuals with and without control

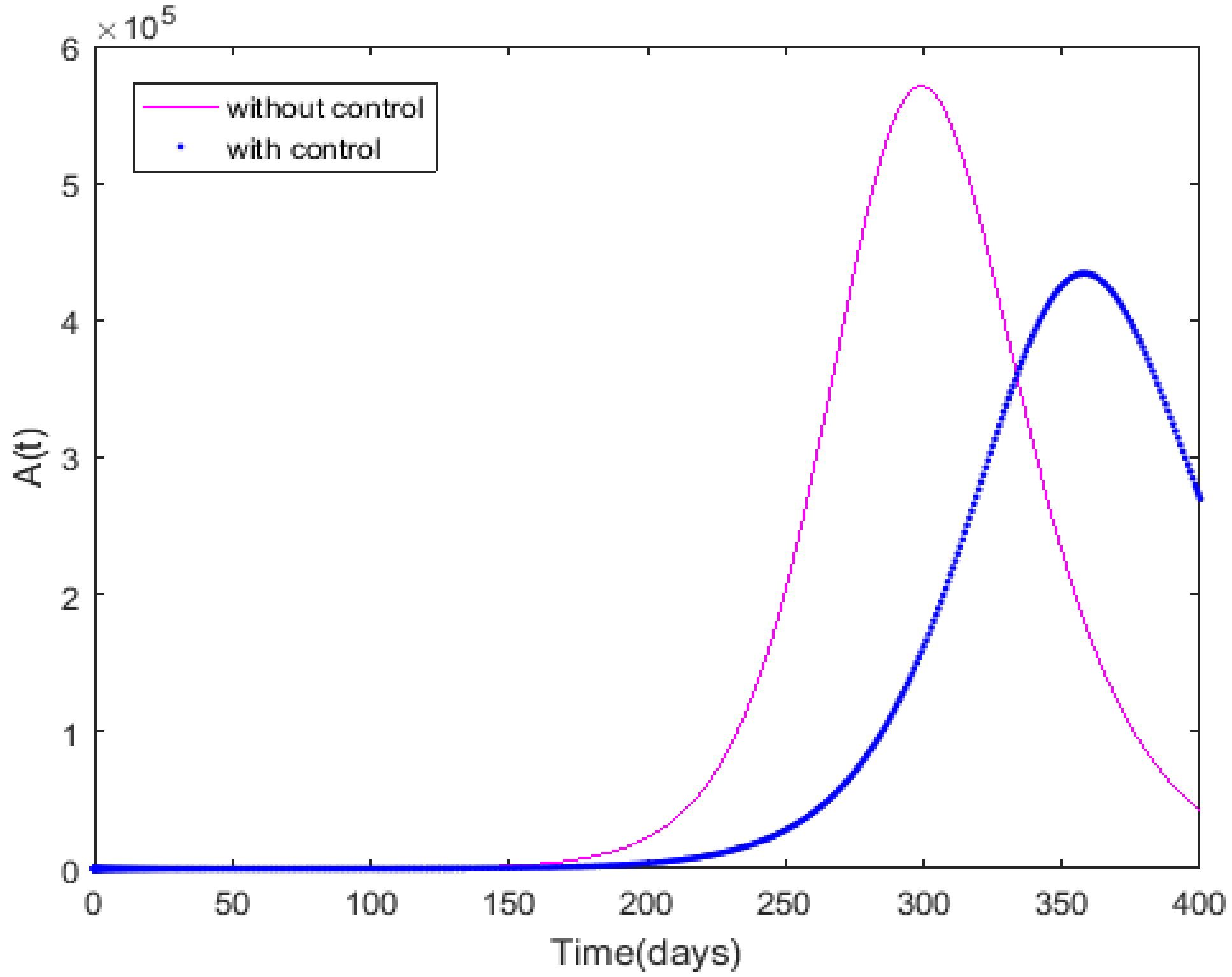

Fig. 13: Behaviour of asymptomatic individuals with and without control

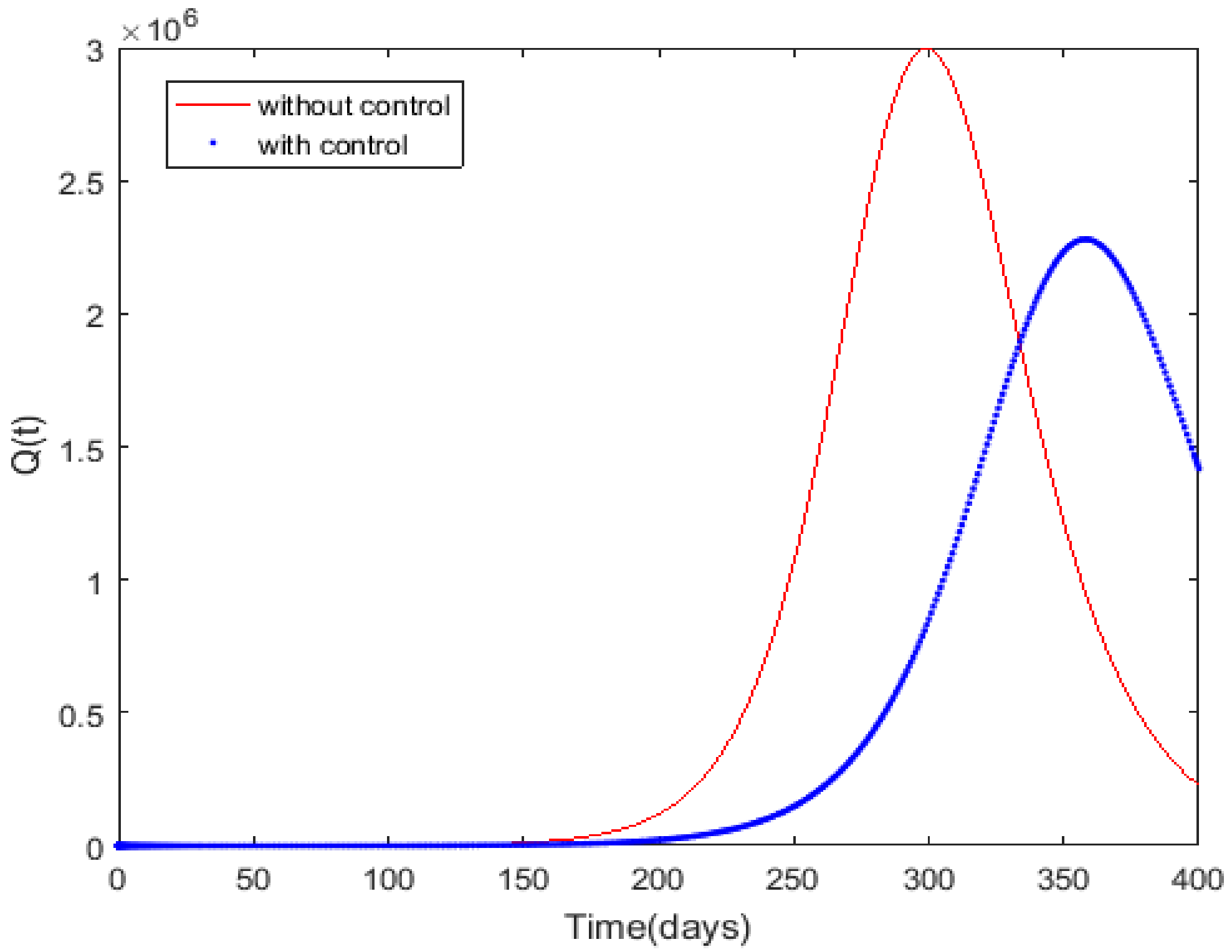

Fig. 14: Behaviour of symptomatic individuals with and without control

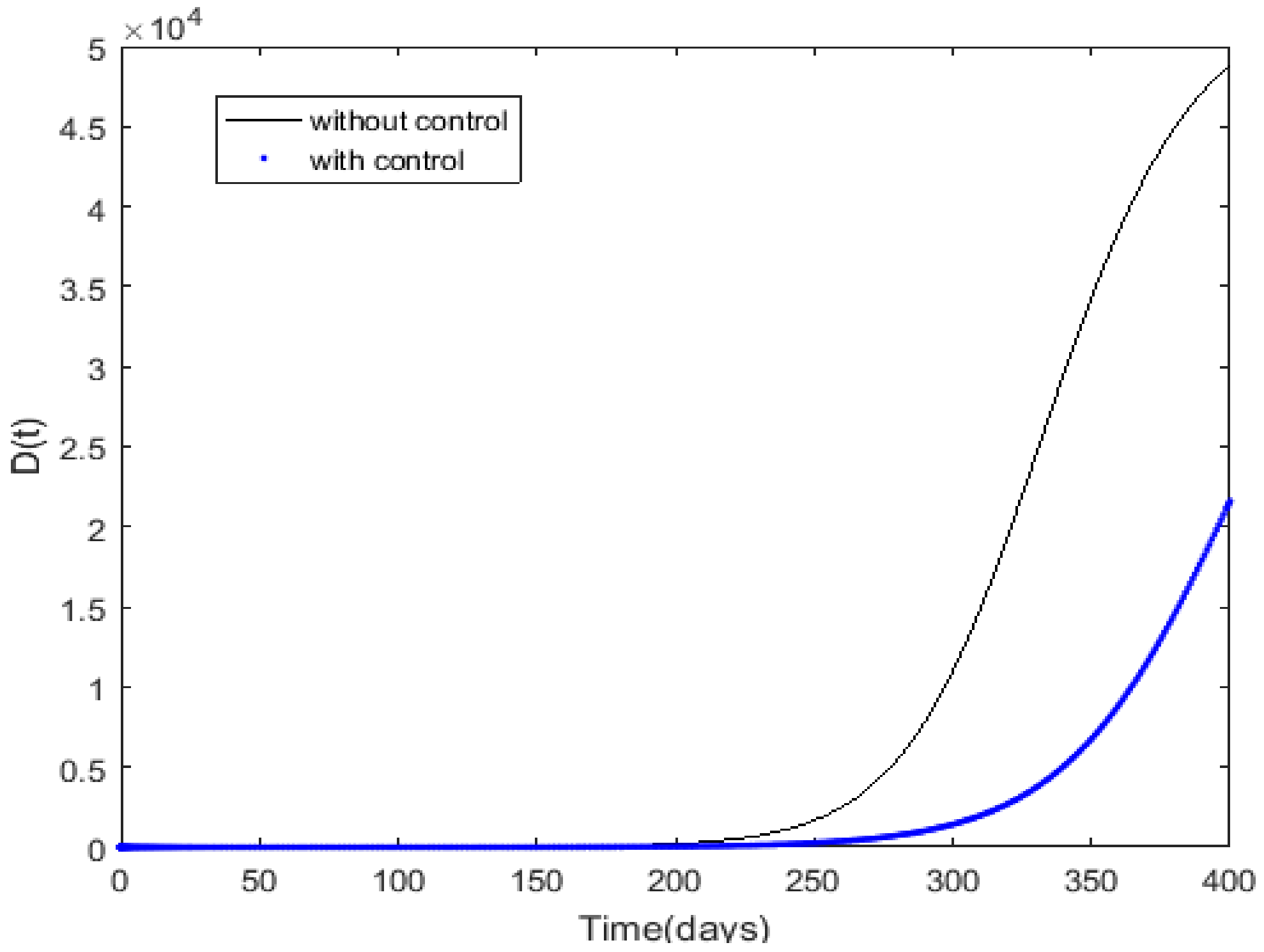

Fig. 15: Behaviour of individuals deceased with and without control

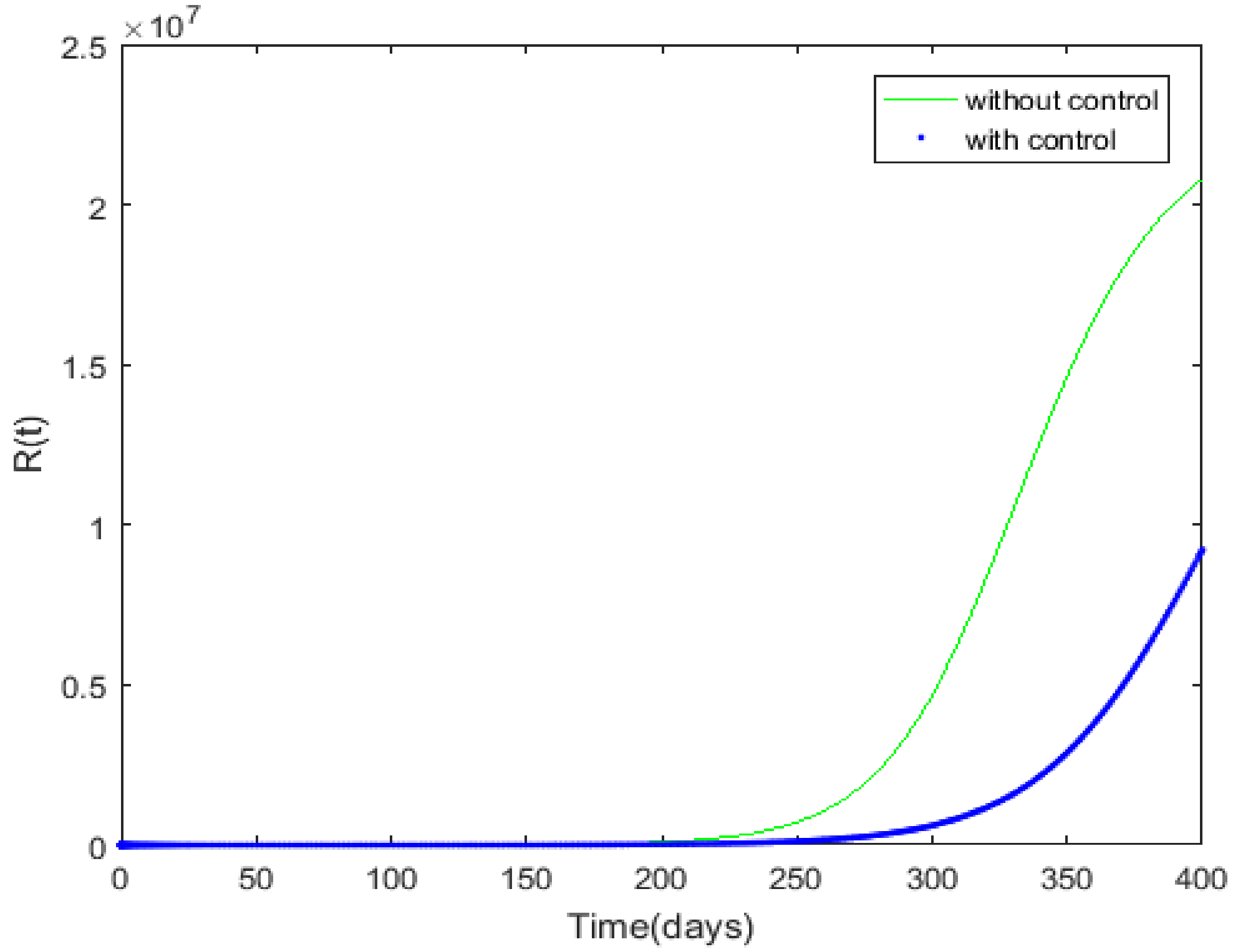

Fig. 16: Behaviour of recovered individuals with and without control

In figure 11, the case with control is higher than a case without control for the susceptible population. Though they both decline, a situation without control decline faster than when there is a control. In figures 12 – 16 there is a decline in the exposed, asymptomatic, symptomatic, deceased and recovered population respectively when there is a control within 400 days. This shows that, fewer infections, recoveries and less people deceased when the nose masks usage are implemented.

We now focus our attention on the vaccination control $u_2$. The goal is to reduce the number of susceptible who contract COVID-19 and increase the number of people vaccinated against the disease. The results are displayed in Figs. 17 – 22.

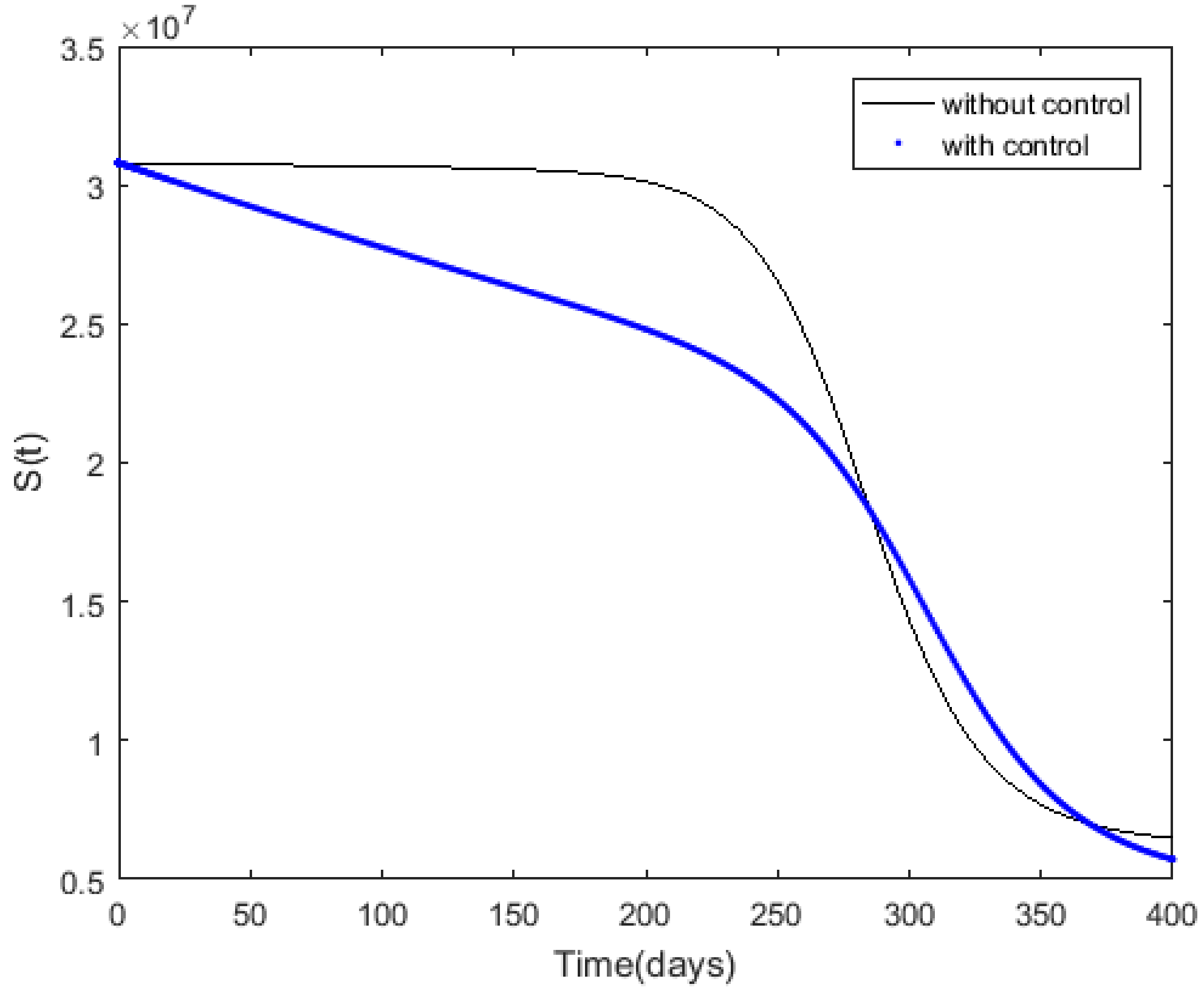

Fig. 17: Behaviour of the Susceptible with and without control

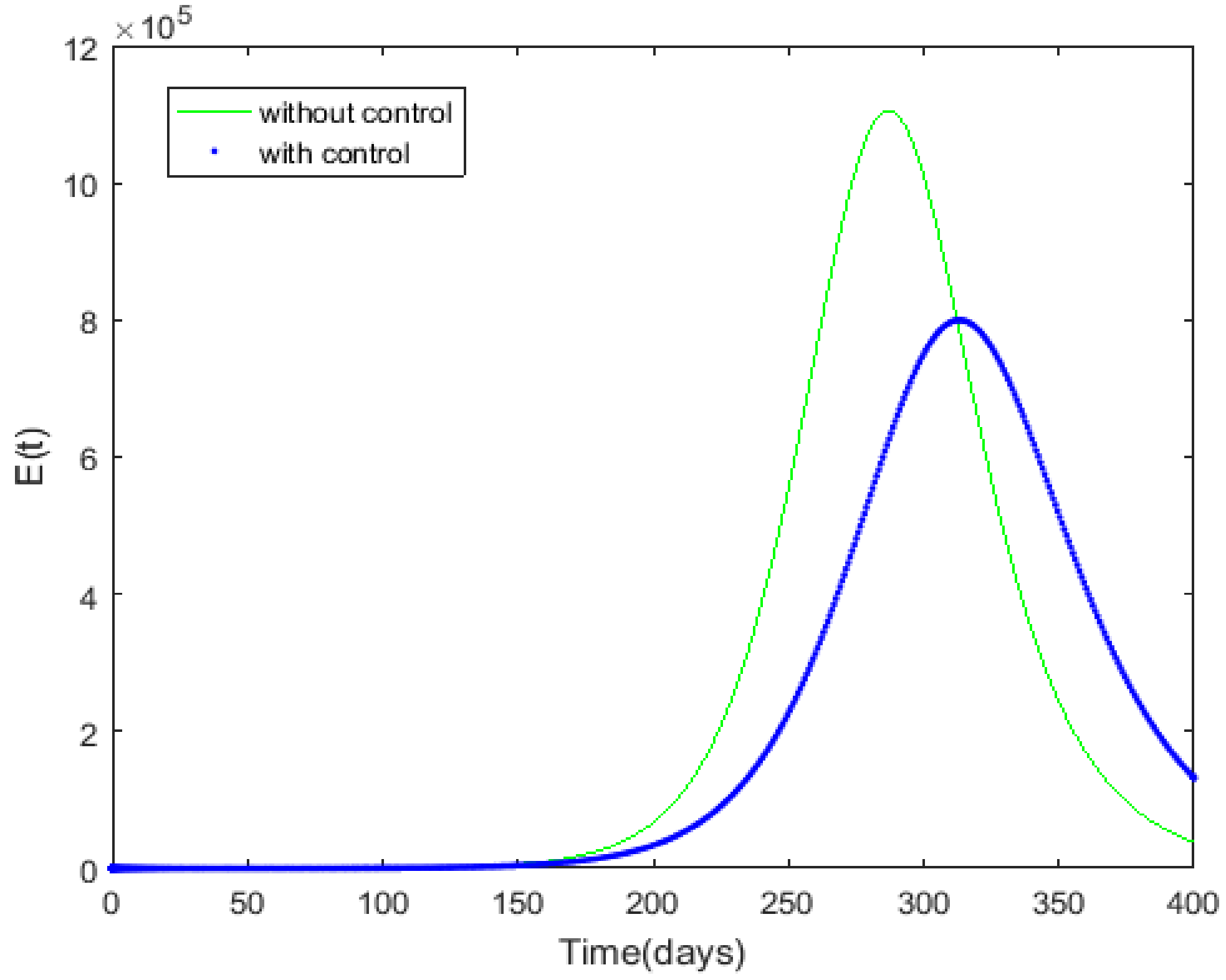

Fig. 18: Behaviour of the exposed individuals with and without control

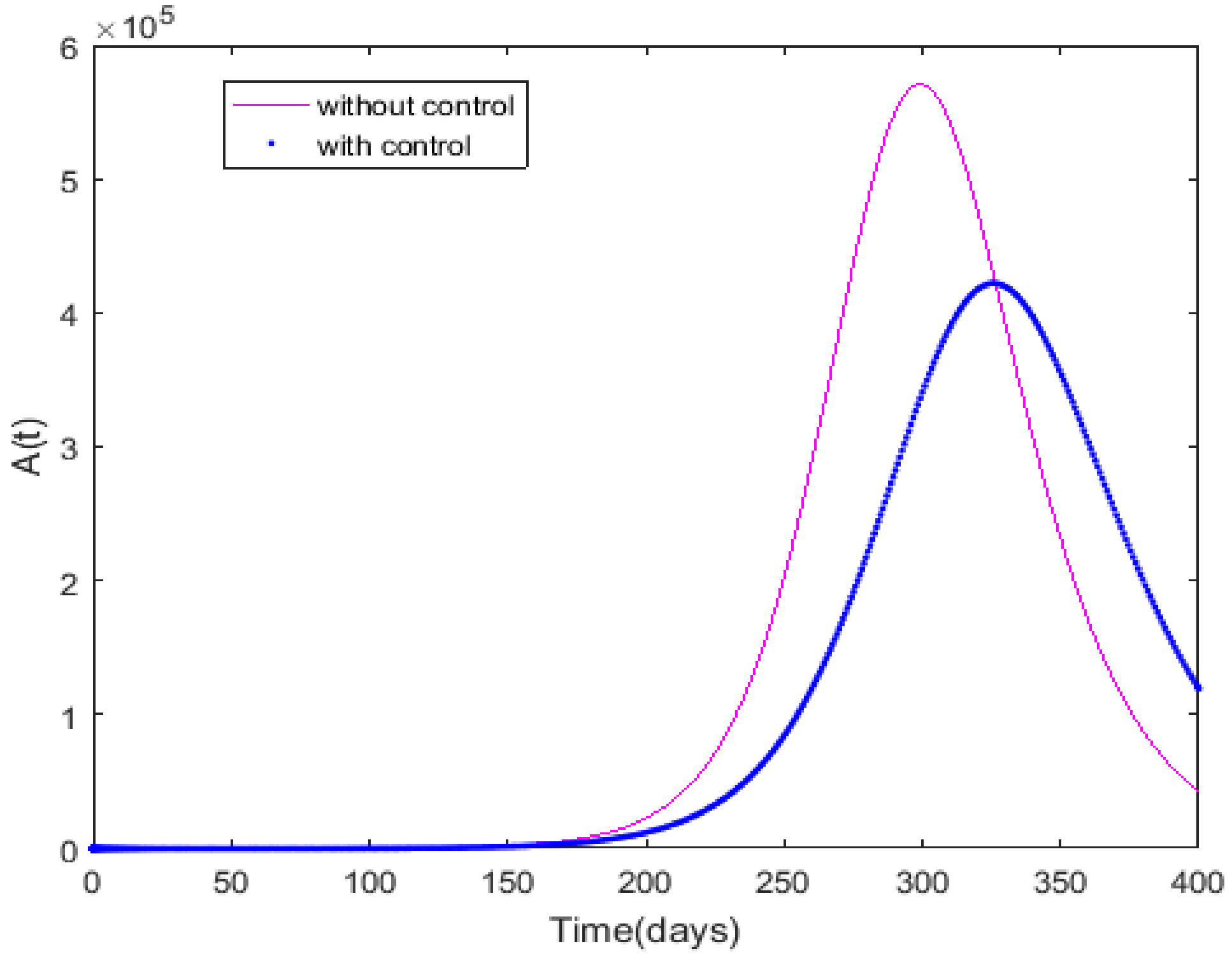

Fig. 19: Behaviour of the asymptomatic with and without control

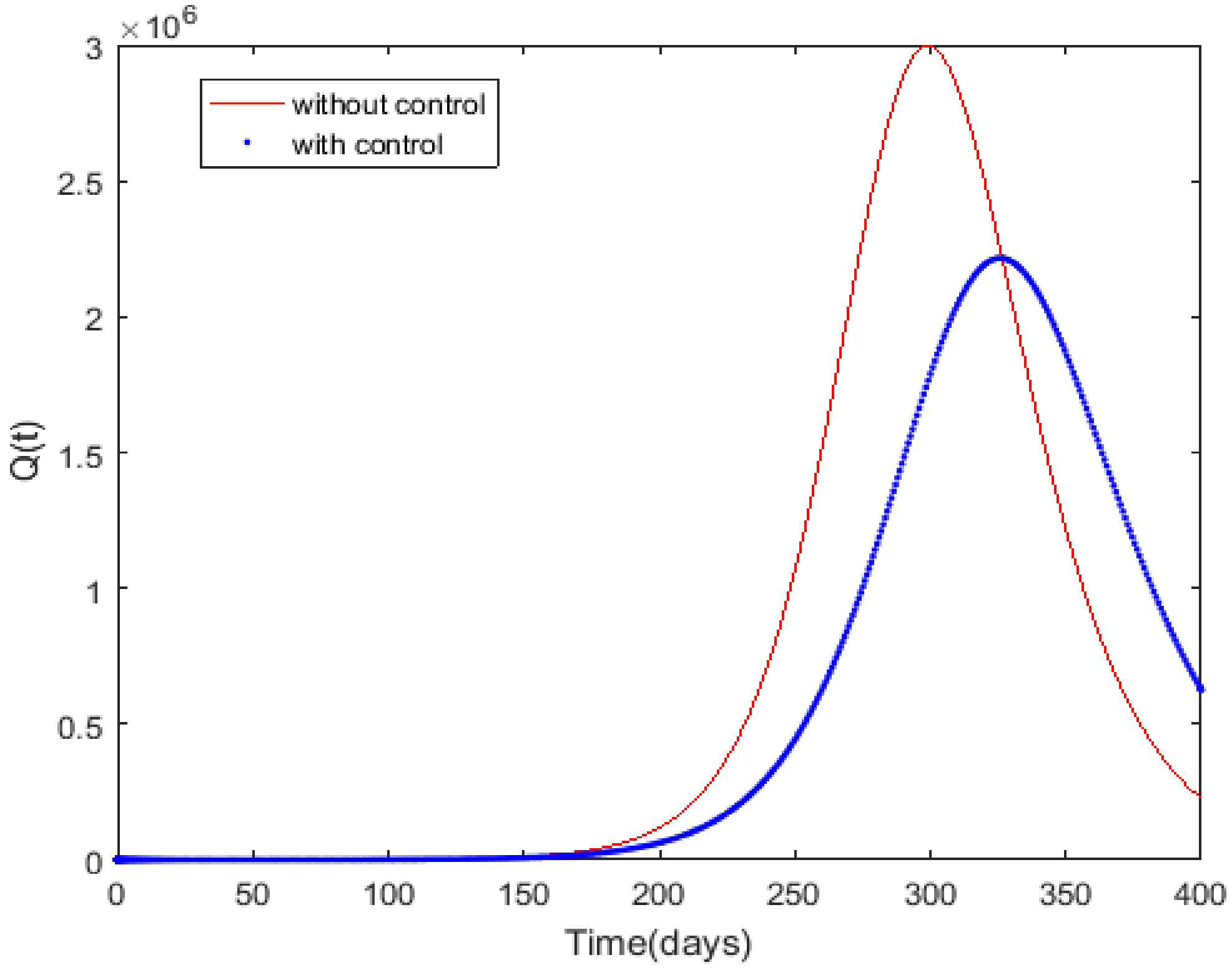

Fig. 20: Behaviour of the symptomatic individuals with and without control

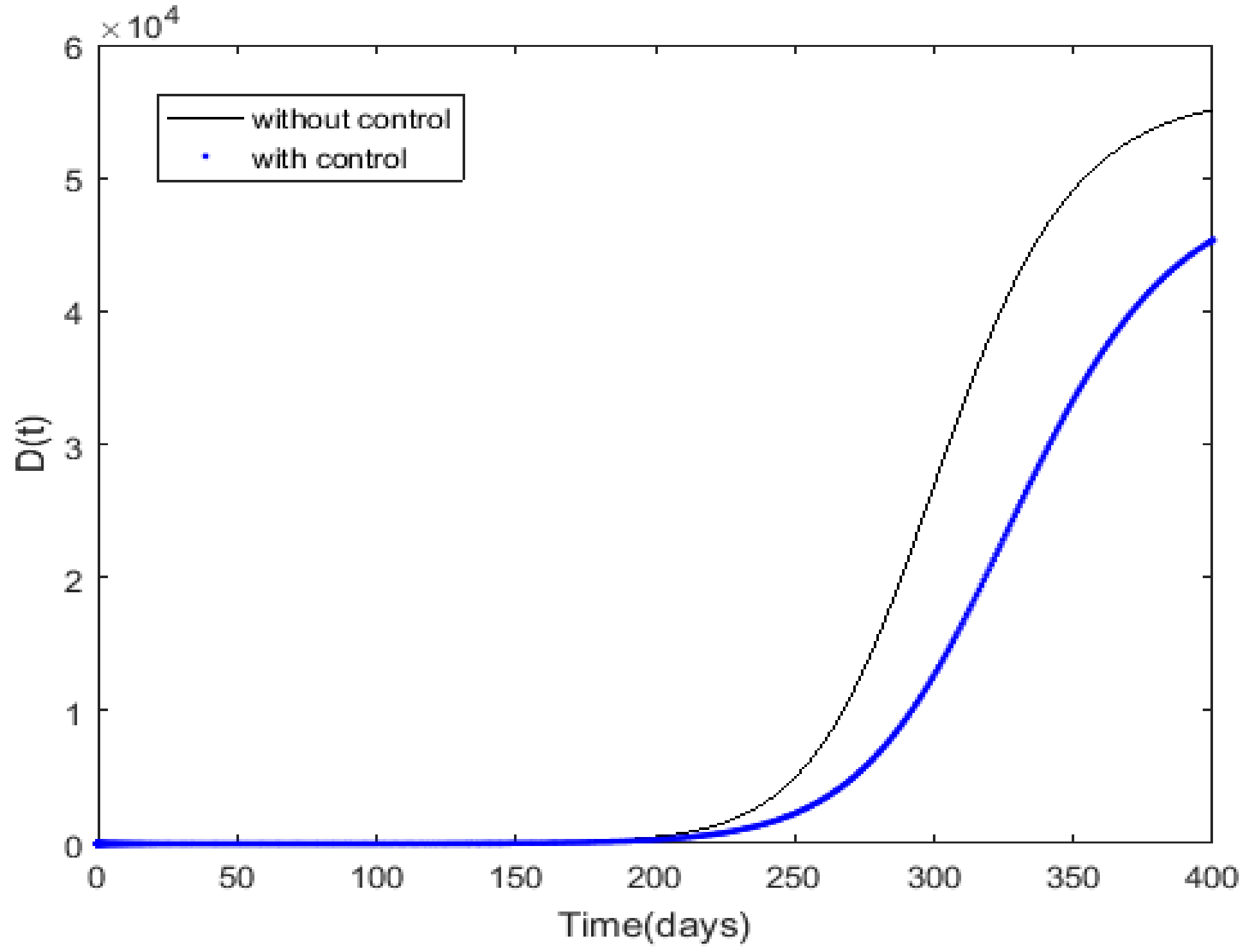

Fig. 21: Behaviour of the deceased compartment with and without control

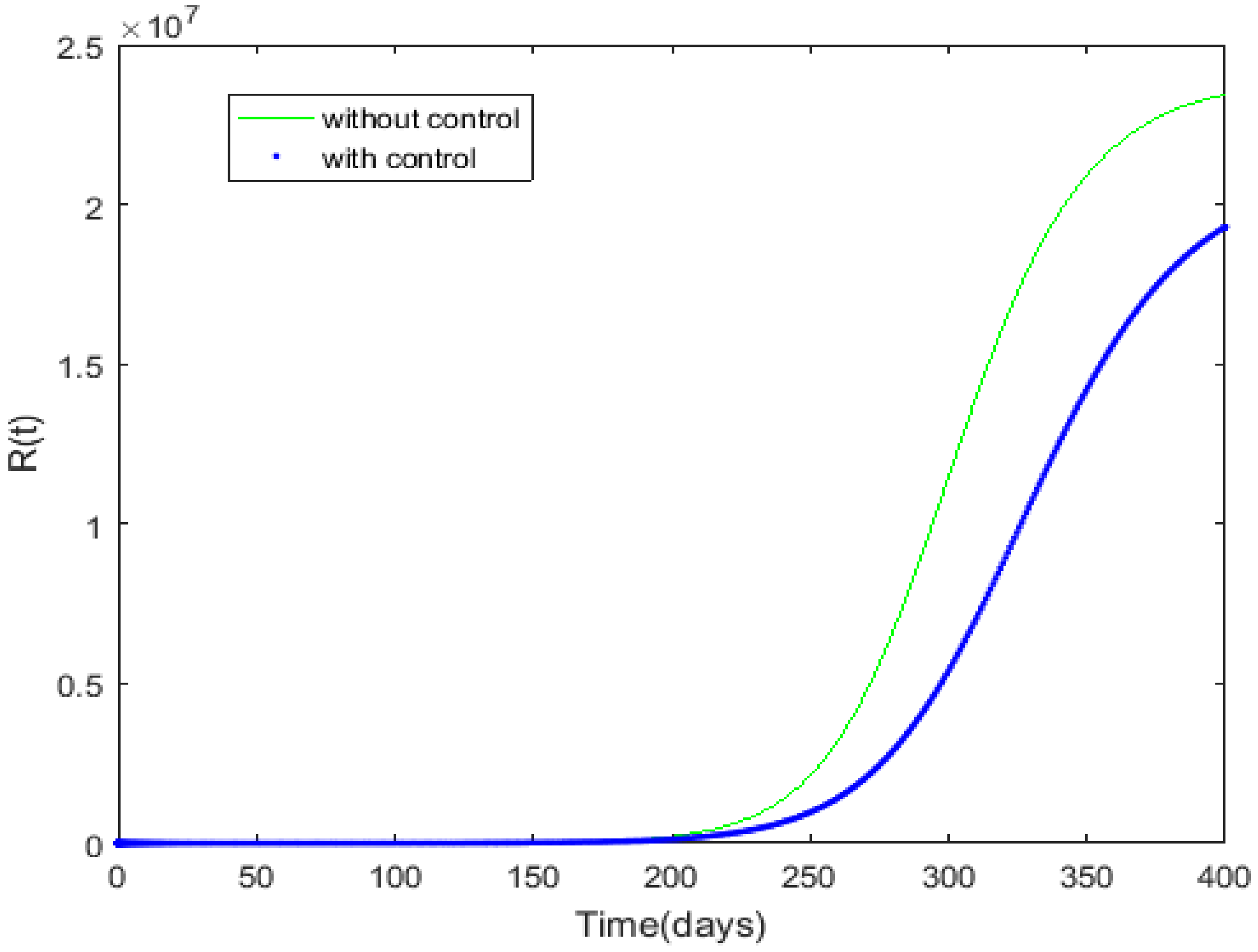

Fig. 22: Behaviour of the recovered class with and without control

In figures 17 – 22 there is a decline in the susceptible, exposed, asymptomatic, symptomatic, deceased and recovered population respectively when vaccination control is implemented within 400 days. This shows that, fewer infections, recoveries and less number of individuals deceased when the vaccination is implemented fully.

## 6 Conclusion

An optimal control model has been formulated to study and control the spread of COVID-19 in Ghana. The basic reproduction, equilibrium points and stability of the equilibrium points has been determined. The model was locally and globally stable for the disease – free equilibrium. The model was validated using COVID-19 data for the period March 2020 to March 2021. The results of the numerical simulation were consistent with the real data from Ghana. The simulation revealed the disease had less impact on the population during the first seven months of the outbreak. Optimal controls were incorporated into the model to determine the effectiveness of two preventive control measures such as the use of a nose mask and vaccination. Both measures were very effective in curtailing the spread of the disease as there is a decline in the number of exposed, asymptomatic, symptomatic and the deceased when the controls were optimized fully. The use of masks though led to a decline in the susceptible population, this was slow as compared to a situation without its usage. The vaccination on the other hand reduces the number of the susceptible individuals faster than a non-vaccination population. This is so because the vaccinated individuals tend to develop immunity to the disease.

**Acknowledgement**

This Manuscript was submitted as a pre-print in the link https://arxiv.org/ftp/arxiv/papers/2201/2201.08224.pdf and has been referenced.

**Data Availability**

The data/information supporting the formulation of the mathematical model in this paper are/is from Ghana health service website: https://www.ghs.gov.gh/covid19/ which has been cited in the manuscript.

**Declaration of conflict of interest**

No conflict of interest regarding the content of this article

Reference

1. A.A. Agyemang K.L. Chin, C.B. Landersdorfer, D. Liew, R. Ofori-Asenso, "Smell and taste Dysfunction in Patients with COVID-19: A systematic Review and Meta - analysis", Mayo Clin. Proc., 2020; 95(8):1621-1631. Doi:10.1016/j.mayocp.2020.05.030
2. A.M Ajbar, E. Ali, A. Ajbar, "Modeling the evolution of the coronavirus disease (COVID-19) in Saudi Arabia",J infect Dev Ctries, 2021; 15(7):918-924, doi:10.3855/jidc.13568.PMID:34343116
3. C. Castillo-Chavez, S. Blower, P. van den Driessche, D. Kirschener, Y. Abdul-Aziz, Mathematical approach for emerging and reemerging infectious diseases. Springer Verlag, New York;
4. D.P. Oran E.J. Topol, "The proportion of SARS-CoV-2 infections that are asymptomatic: A systematic review", Annals of internal medicine, 2021; 174 (5): M20-6976, doi: 107326/M20-6976
5. European Centre for Disease Prevention and Control, "Transmission of COVID-19". https://www.ecdc.europa.eu/en/covid-19/latest-evidence/transmission, Retrieved on 12th September, 2020
6. I. Appiah-Otoo, M.B. Kursah, "Modelling spatial variations of novel coronavirus disease (COVID-19): evidence from a global perspective", GeoJournal,Vol. 24: 1-15;2021, doi:10.1007/s10708-021-10427-0
7. J. Page, D. Hinshaw, B. Mckay,"In the hunt for covid-19origin, patient zero points to the second Wuhan Market – The man with the first confirmed infection of the new coronavirus told the WHO team that his parents had shopped there", The wall street Journal, 2021: (Retrieve 10th November, 2021)

8. J. Saniasiaya and M.A. Islam, “Prevalence and characteristics of taste disorders in Cases of COVID-19: Systematic review and meta-analysis of 29349 patients”, Otolaryngology – Head and neck Surgery. 165(1): 33-42. doi:10.1177/0194599820981018.
9. L.S. Pontryagin, V.G. Boltyanskii, R.V. Gamkrelidze, and E.F. Mishchenko, The mathematical theory of optimal processes 1963, VIII +360S. New York/London. Wiley and Sons. https://onlinelibrary.wiley.com/doi/abs/10.1002/zamm.19630431023
10. M. A. Islam, “Prevalence of headache in patients with coronavirus disease, 2019 (COVID-19): Systematic review and meta-analysis of 14275 patients”, Frontiers in Neurology, 2020; 11:562634, doi:10.3389/fneur.2020.562634
11. M. A. Islam, “Prevalence and characteristics of fever in adult and paediatric patients with coronavirus disease 2019 (COVID-19): Systematic review and meta-analysis of 17515 patients”, PLOS ONE, 2021; 16:4, doi:10.1371/fneur.2020.562634
12. M. Veera Krishna, “Mathematical modeling of diffusion and control of COVID-19”, Infectious Disease Modeling, Vol. 5, 2020; pg 588 – 597, ISSN 2468-0427, ISSN 2468-0427, https://doi.org/10.1016/j.idm.2020.08.009
13. M. Veera Krishna, J. Prakash, “Mathematical modeling on phase-based transmissibility of coronavirus”, Infectious Didease Modelling , Vol. 5, 2020; pg 375 – 385, ISSN 2468-0427, https://doi.org/10.1016/j.idm.2020.06.005
14. U.S. Centers for Disease Control and Prevention (CDC),”Interim clinical guidance for the management of patients with confirmed coronavirus disease (COVID-19)”, 2020; retrieved on 5th July, 2021
15. U.S. Centers for Disease Control and Prevention, "How COVID-19 Spreads", *2020;* https://www.cdc.gov/coronavirus/2019-ncov/prevent-getting-sick/how-covid-spread.html. Retrieved on 20th November, 2020.
16. S.P. Dharmaratne, S. Sudaraka, I. Abeyagunawardena, G. Wasantha, “Estimation of the basic reproduction number (R0) for the novel coronavirus disease in Sri Lanka”, Virol J 17, 144 (2020). https://doi.org/10.1186/s12985-020-01411-0
17. World Health Organization (WHO). Coronavirus disease (COVID-19). 2021 https://www.who.int/westernpacific/health-topics/detail/coronavirus
18. Worldometers, “COVID-19 Coronavirus Pandemic”, 2021, http://www.worldometers.info/coronavirus
19. S.P. Dharmaratne, S. Sudaraka, I. Abeyagunawardena, G. Wasantha, “Estimation of the basic reproduction number (R0) for the novel coronavirus disease in Sri Lanka”, Virol J 17, 144 (2020). https://doi.org/10.1186/s12985-020-01411-0
20. C. Castillo-Chavez, S. Blower, P. van den Driessche, D. Kirschener, Y. Abdul-Aziz, Mathematical approach for emerging and reemerging infectious diseases. Springer Verlag, New York;
21. Ghana Birth rate; 2021, www.indexmundi.com retrieved on 11th December, 2021
22. I. Ahmed, G. U. Modu, A. Yusuf, P. Kumam, and I. Yusuf, A mathematical model of coronavirus disease (COVID-19) containing asymptomatic and symptomatic classes, Elsevier public health emergency collection, Results Phys. 2021; 21:103776, doi: 10.1016/j.rinp.2020.103776
23. L.S. Pontryagin, V.G. Boltyanskii, R.V. Gamkrelidze, and E.F. Mishchenko, The mathematical theory of optimal processes 1963, VIII +360S. New York/London. Wiley and Sons. https://onlinelibrary.wiley.com/doi/abs/10.1002/zamm.19630431023
24. Ghana Health Service, ‘COVID-19 Updates|Ghana’, https://www.ghs.gov.gh/covid19/,

(Retrieved on 1st December, 2021)
25. J. Duncan, "Two cases of coronavirus confirmed in Ghana". Citi Newsroom https://citinewsroom.com/2020/03/two-cases-of-coronavirus-confirmed-in-ghana/ (Retrieved 20th March 2021).
26. S. Okyere, J.A. Prah and A.N. Owusu-Sarpong, "An Optimal Control model of the transmission dynamics of SARS – CoV – 2 (COVID-19) in Ghana", arXiv:2201.08224 **[q-bio.PE]** https://arxiv.org/ftp/arxiv/papers/2201/2201.08224.pdf
27. The Conversation, 'Ghana's debt makes development impossible: here are the solutions', 2022; https://theconversation.com/ghanas-debt-makes-development-impossible-here-are-some-solutions-176580#:~:text=That's%20close%20to%20the%20level,improving%20at%2045%25%20in%202022. (Retrieved online 5th March, 2022)
28. D.T. Aduhene., and E. Osei-Assibey, E., 'Socio-economic impact of COVID-19 on Ghana's economy: challenges and prospects', *International Journal of Social Economics*, (2021) Vol. 48 No. 4, pp. 543-556. https://doi.org/10.1108/IJSE-08-2020-0582
29. F. Ndaïrou, I. Area, J. J. Nieto, D. F.M. Torres, "Mathematical modeling of COVID-19 transmission dynamics with a case study of Wuhan", Chaos, Solitons & Fractals, Volume 135, 2020, 109846, ISSN 0960-0779, https://doi.org/10.1016/j.chaos.2020.109846
30. J.Y.T Mugisha, J. Ssebuliba, J.N. Nakakawa, C.R. Kikawa, A. Ssematimba, "Mathematical modeling of COVID-19 transmission dynamics in Uganda: Implications of complacency and early easing of lockdown", PLoS ONE, 2021; 16(2): e0247456. https://doi.org/10.1371/journal.pone.0247456
31. S. M. Garba and J. M.-S. Lubuma and B. Tsanou, 'Modeling the transmission dynamics of the COVID-19 Pandemic in South Africa', Mathematical Biosciences, 2020; 328 (108441): 0025-5564, https://doi.org/10.1016/j.mbs.2020.108441
32. X. Fu, Q. Ying, T. Zeng, T. Long, and Y. Wang, "Simulating and forecasting the cumulative confirmed cases of SARS-CoV-2 in China by Boltzmann function-based regression analyses," *Journal of Infection*, 2020; vol. 80, no. 5, pp. 578–606, https://doi.org/10.1016/j.jinf.2020.02.019
33. S. Nana-Kyere, F.A Boateng, P. Jonathan, A. Donkor, G. K. Hoggar, B. D. Titus D. Kwarteng, and I. K. Adu, 'Global Analysis and optimal control model of COVID-19', Computational and Mathematical Methods in Medicine, 2022; vol. 2022, Article ID 9491847, 20 pages, https://doi.org/10.1155/2022/9491847, https://www.hindawi.com/journals/cmmm/2022/9491847/
34. D. Dwomoh, S. Iddi, B. Adu, J. M. Aheto, K. M. Sedzro, Julius Fobil, Samuel Bosomprah, 'Mathematical modeling of COVID-19 infection dynamics in Ghana: Impact evaluation of integrated government and individual level interventions', Infectious Disease Modelling, Volume 6, 2021, Pages 381-397, ISSN 2468-0427, https://doi.org/10.1016/j.idm.2021.01.008.
35. F. Ndaïrou, I. Area, J. J. Nieto, D. F.M. Torres, "Mathematical modeling of COVID-19 transmission dynamics with a case study of Wuhan", Chaos, Solitons & Fractals, Volume 135, 2020, 109846, ISSN 0960-0779, https://doi.org/10.1016/j.chaos.2020.109846

36. D. D. Sosu, ‘Cumulative number of COVID-19 vaccine doses administered in Ghana 2022’, Statista, 2022, https://statista.com/statistics/1227780/total-number-of-covid-19-vaccination-doses-in-ghana/ (Retrieved online 1st April, 2022)
37. M. A. Khan, K. Ali, E. Bonyah, K. O. Okosun, S. Islam, and A. Khan, “Mathematical modeling and stability analysis of pine wilt disease with optimal control,” *Scientific Reports*, vol. 7, no. 1, 2017, https://www.nature.com/articles/s41598-017-03179-w
38. P. Bogacki and L. F. Shampine, “An efficient Runge-Kutta (4,5) pair,” *Computers & Mathematics with Applications*, vol. 32, no. 6, pp. 15–28, 1996. https://www.sciencedirect.com/science/article/pii/0898122196001411?via%3Dihub
39. L. F. Shampine, “Some practical Runge-Kutta formulas,” *Mathematics of Computation*, vol. 46, no. 173, pp. 135–150, 1986. https://www.ams.org/journals/mcom/1986-46-173/S0025-5718-1986-0815836-3/
40. E. Jung, S. Lenhart, and Z. Feng, “Optimal control of treatments in a two-strain tuberculosis model,” Discrete & Continuous Dynamical Systems-*B*, vol. 2, no. 4, pp. 473–482, 2002. https://www.aimsciences.org/article/doi/10.3934/dcdsb.2002.2.473
41. Cristiana J. Silva, Helmut Maurer, Delfim F. M. Torres. Optimal control of a Tuberculosis model with state and control delays. *Mathematical Biosciences & Engineering*, 2017, 14 (1): 321-337. doi: 10.3934/mbe.2017021. https://www.aimsciences.org/article/doi/10.3934/mbe.2017021?utm_source=TrendMD&utm_medium=cpc&utm_campaign=Mathematical_Biosciences_%2526_Engineering_TrendMD_0
42. C. J. Silva, D. F. M. Torres, ‘Optimal control strategies for tuberculosis treatment: A case study in Angola’, *Numerical Algebra, Control and Optimization*, 2012, 2 (3) : 601-617. doi: 10.3934/naco.2012.2.601. http://www.aimsciences.org/journals/displayArticlesnew.jsp?paperID=7529&utm_sour
43. C. J. Silva, D. F. M. Torres, ‘A TB-HIV/AIDS coinfection model and optimal control treatment. *Discrete and Continuous Dynamical Systems*, 2015, 35 (9) : 4639-4663. doi: 10.3934/dcds.2015.35.4639
44. G.W. Swan, ‘An optimal control model of diabetes mellitus’, Bulletin of Mathematical Biology, Volume 44, Issue 6, 1982, Pages 793-808, ISSN 0092-8240, https://doi.org/10.1016/S0092-8240(82)80043-8.
45. S. Nana-Kyere, F.A Boateng, P. Jonathan, A. Donkor, G. K. Hoggar, B. D. Titus D. Kwarteng, and I. K. Adu, ‘Global Analysis and optimal control model of COVID-19’, Computational and Mathematical Methods in Medicine, 2022; vol. 2022, Article ID 9491847, 20 pages, https://doi.org/10.1155/2022/9491847, https://www.hindawi.com/journals/cmmm/2022/9491847/
46. S. İ. Araz, Analysis of a Covid-19 model: Optimal control, stability and simulations, Alexandria Engineering Journal, Volume 60, Issue 1, 2021, Pages 647-658, ISSN 1110-0168, https://doi.org/10.1016/j.aej.2020.09.058.
47. C. T. Deressa and G. F. Duressa, “Modeling and optimal control analysis of transmission dynamics of COVID-19: the case of Ethiopia,” *Alexandria Engineering Journal*, vol. 60, no. 1, pp. 719–732, 2021. https://www.sciencedirect.com/science/article/pii/S1110016820305202?via%3Dihub
48. N. R. Sasmita, M. Ikhwan, S. Suyanto, and V. Chongsuvivatwong, “Optimal control on a mathematical model to pattern the progression of coronavirus disease 2019 (COVID-19) in Indonesia,” *Global Health Research and Policy*, vol. 5, no. 1, pp. 1–12, 2020.
49. A. Perkins and G. España, “Optimal control of the COVID-19 pandemic with nonpharmaceutical interventions,” *Bulletin of Mathematical Biology*, vol. 82, no. 9, p. 118, 2020.

50. M.L. Diagne, H. Rwezaura, S.Y. Tchoumi, J.M. Tchueche, ‘A Mathematical Model of the COVID-19 with vaccination and treatment’, Computational and Mathematical Methods in Medicine, 2021, vol. 2021, Article ID 1250129, 16 pages. https://www.hindawi.com/journals/cmmm/2021/1250129/
51. I. Ahmed, G. U. Modu, A. Yusuf, P. Kumam and I. Yusuf, A mathematical model of coronavirus disease (COVID-19) containing asymptomatic and symptomatic classes, Elsevier public health emergency collection, Results Phys. 2021; 21:103776, doi: 10.1016/j.rinp.2020.103776. https://www.ncbi.nlm.nih.gov/pmc/articles/PMC7787076/
52. The World Bank ‘Life expectancy at birth’, https://data.worldbank.org/indicator/SP.DYN.LE00.IN?locations=GH
53. Ghana Statistical Service, ‘The 2021 population and housing census’, 2021, https://census2021.statsghana.gov.gh/index.php (Retrieved online 5th March, 2022)
54. E. Bonyah, A. K. Sagoe, D. Kumar, S. Deniz, ‘Fractional optimal control dynamics of coronavirus model with Mittag–Leffler law, Ecological Complexity, 2021; 45(100880): 1476-945X, https://doi.org/10.1016/j.ecocom.2020.100880.